\documentclass[12pt]{article}
\pdfoutput=1

\usepackage{draft}

\begin{document}

\begin{titlepage}

\preprint{CALT-TH-2021-006}

\begin{center}

\hfill \\
\hfill \\
\vskip 1cm

\title{
Topological Field Theory with Haagerup Symmetry
}

\author{Tzu-Chen Huang$^{a}$, Ying-Hsuan Lin$^{b,a}$}

\address{${}^a$Walter Burke Institute for Theoretical Physics,
\\ 
California Institute of Technology, Pasadena, CA 91125, USA}

\address{${}^b$Jefferson Physical Laboratory, Harvard University, Cambridge, MA 02138, USA}

\email{jimmy@caltech.edu, yhlin@fas.harvard.edu}

\end{center}

\vfill

\begin{abstract}
We construct a (1+1)$d$ topological field theory (TFT) whose topological defect lines (TDLs) realize the transparent Haagerup $\mathcal{H}_3$ fusion category.  This TFT has six vacua, and each of the three non-invertible simple TDLs hosts three defect operators, giving rise to a total of 15 point-like operators.  The TFT data, including the three-point coefficients and lasso diagrams, are determined by solving all the sphere four-point crossing equations and torus one-point modular invariance equations.  We further verify that the Cardy states furnish a non-negative integer matrix representation under TDL fusion.  While many of the constraints we derive are not limited to the this particular TFT with six vacua, we leave open the construction of TFTs with two or four vacua.  Finally, TFTs realizing the Haagerup $\mathcal{H}_1$ and $\mathcal{H}_2$ fusion categories can be obtained by gauging algebra objects.  This note makes a modest offering in our pursuit of exotica and the quest for their eventual conformity.
\end{abstract}

\vfill

\end{titlepage}

\tableofcontents

\section{Introduction}
\label{Sec:Intro}

The best cultivated terrains in the landscape of (1+1)$d$ conformal field theories (CFTs) are rational conformal field theories (RCFTs) \cite{Friedan:1983xq}, free theories, and orbifolds \cite{Dixon:1985jw,Dixon:1986jc} thereof.  Exactly marginal deformations of orbifold twist fields bring us into more interesting realms, and when roamed far enough provide candidates with weakly coupled holographic duals.  But the full landscape is believed to be vaster.  The conformal bootstrap bounds on various quantities such as the twist gap \cite{Collier:2016cls,Benjamin:2019stq,Benjamin:2020swg} are not saturated by known CFTs, and numerical studies of certain renormalization group flows, such as that from the three-coupled three-state Potts model \cite{Dotsenko:1999erv}, indicate the existence of fixed points with irrational central charges.  However, such fixed points are evasive of current analytic methods.  Even for RCFTs, a full classification has not been achieved.

The full set of interesting observables in a (1+1)$d$ CFT is not limited to the correlation functions of local operators.  There are boundaries and defects that interact with the local operators in nontrivial ways, and are together subject to stringent consistency conditions.  Some of the data, like the fusion category \cite{Etingof:aa,etingof2016tensor} furnished by the topological defect lines (TDLs) \cite{Bhardwaj:2017xup,Tachikawa:2017gyf,Chang:2018iay}, are mathematically rigid structures that exist independently of quantum field theory.  A simple example of a fusion category is a group-like category, which consists of the specification of a discrete symmetry group together with its anomaly.  Fusion categories generalize symmetries and anomalies, and constrain the deformation space of quantum field theory.  The preceding remarks beg the following question:

{\bf Q1:} {\it Given a fusion category, is there a (1+1)$d$ CFT whose TDLs (or a subset thereof) realize the said category?}

The (2+1)$d$ Turaev-Viro theory \cite{Turaev:1992hq} or Levin-Wen string-net model \cite{Levin:2004mi} constructed out of a fusion category $\cC$ is a bulk phase whose anyons are described by the Drinfeld center $\cZ(\cC)$. By placing the bulk phase on a slab between a gapped boundary and another boundary condition $B$, and further compactifying on a circle, the resulting theory $T[B]$ would be a CFT with TDLs described by $\cC$. {\bf Q1} is thus interpreted as the existence/classification problem of boundary conditions for the bulk phase.\footnote{The authors thank Shu-Heng Shao and Yifan Wang for discussions. 
}
From a purely (1+1)$d$ perspective, statistical height models which take $\cC$ (and the choice of a distinguished object) as the microscopic input have recently been shown by Aasen, Fendley, and Mong \cite{Aasen:2020jwb} to host macroscopic TDLs described by $\cC$.  One can explore the phases of such models in search of a CFT.

A closely related question is the following:

{\bf Q2:} {\it Given a modular tensor category (MTC), is there a vertex operator algebra (VOA) whose representations realize the said category?}

The phrase VOA could be replaced by diagonal RCFT, in which the fusion ring of Verlinde lines (TDLs commuting with the VOA) is isomorphic to the fusion ring of the VOA representations.  The correspondence between MTC and (1+1)$d$ RCFT traces its origin to a seminal series of papers by Moore and Seiberg \cite{Moore:1988qv,Moore:1988ss,Moore:1988uz,Moore:1989vd,Moore:1989yh}, and is conjectured to be one-to-one thought a construction or proof is lacking.  Note that an affirmative answer to {\bf Q2} implies an affirmative answer to {\bf Q1}:  Given a fusion category $\cC$, if one can find a VOA that realizes the Drinfeld center $\cD(\cC)$, then gauging the diagonal RCFT by an algebra object gives a non-diagonal RCFT whose TDLs realize $\cC \boxtimes \cC^{\rm op}$.\footnote{The authors thank Sahand Seifnashri for a discussion.
}

The explicit realization of many categories in CFT is not known.  A famous example is the Haagerup fusion category, which has a special place in the history of category and subfactor theory.  Subfactors have inherent categorical structure, and serve as a major source of fusion categories.  While Ocneanu \cite{ocneanu1988quantized} and Popa \cite{popa1994classification} classified subfactors with Jones indices less than or equal to 4, Haagerup and Asaeda \cite{asaeda1999exotic} constructed one---the Haagerup subfactor---with Jones index ${5+\sqrt{13}\over2}$, the smallest above 4 \cite{haagerup1994principal}.  As the title of \cite{asaeda1999exotic} suggests, the Haagerup subfactor was deemed {\it exotic} since its construction at the time did not fit into any infinite family.  Later work by Izumi \cite{izumi2001structure}, Evans and Gannon \cite{Evans:2010yr} postulated that the Haagerup subfactor does in fact fit into an infinite family, and furthermore constructed the first few members.  This development suggested that the Haagerup may not be exotic after all.  Nonetheless, for various categorical conjectures, the explicit demonstration in the case of Haagerup is viewed as a key test of a conjecture's legitimacy and generality.

There are actually three inequivalent unitary Haagerup fusion categories, commonly denoted by $\cH_1, \, \cH_2, \, \cH_3$.  Most of this note concerns the Haagerup $\cH_3$ fusion category, which did not descend directly from the Haagerup subfactor of \cite{asaeda1999exotic,haagerup1994principal}, but was instead constructed by Grossman and Snyder \cite{Grossman_2012}.  Because the fusion ring (reviewed in Section~\ref{Sec:Fusion}) is non-commutative, the Haagerup $\cH_3$ fusion category cannot possibly be realized by Verlinde lines \cite{Verlinde:1988sn,Petkova:2000ip,Drukker:2010jp,Gaiotto:2014lma} in a diagonal RCFT.  To our knowledge, its realization as general TDLs (need not commute with the full VOA) is not known in any CFT.  To connect to Verlinde lines, one must consider the MTC that is Drinfeld center of Haagerup.  In fact, Evans and Gannon \cite{Evans:2010yr} constructed $c = 8$ characters for the Haagerup modular data, and used it to surmise possible constructions of the VOA through the Goddard-Kent-Olive coset construction \cite{Goddard:1986ee} and its generalizations \cite{dong1999coset,Li:2000it,Gaberdiel:2016zke}, or through the generalized orbifold construction (gauging an algebra object) of Carqueville, Fr\"ohlich, Fuchs, Runkel, and Schweigert \cite{Fuchs:2002cm,Frohlich:2009gb,Carqueville:2012dk} (see \cite{Bhardwaj:2017xup} for a recent discussion).  Recently, Wolf \cite{Wolf:2021kkq} considered the Haagerup anyon chain and numerically searched for CFT phases, but with inconclusive results.\footnote{Anyon chains generalize the golden chain of Feiguin et al \cite{Feiguin:2006ydp}, and arise in a limit of the statistical height models of Aasen, Fendley, and Mong \cite{Aasen:2020jwb}.
}
To date, a {\it bona fide} construction remains an important open question.  By trying to construct CFTs that realize more exotic fusion categories, one hope is that light would be shed beyond the current borders of known (R)CFTs.

Concerning the gapped phases of (1+1)$d$ quantum field theory, described by (1+1)$d$ topological field theories (TFTs) extended by defects \cite{Runkel:2008gr,Davydov:2011kb}, a related but simpler question can be asked:\footnote{There are various notions of TFT with different amounts of structure, the most common being closed TFT \cite{dijkgraaf1989geometrical,Sawin:1995rh,Abrams:1996ty,kock2004frobenius} and open/closed TFT \cite{Alexeevski:2002rp,Lauda:2005wn,Lazaroiu:2000rk,Moore:2006dw}.  The defect TFT of \cite{Runkel:2008gr,Davydov:2011kb} is an overarching formalism that can incorporate multiple closed TFTs and their boundaries and interfaces.  The minimal structure that incorporates the data of TDLs is a defect TFT containing a single closed TFT; mathematically speaking, it is a bicategory with a single object, whose 1-morphisms are the TDLs, and whose 2-morphisms are the local and defect operators.  The full enrichment by boundaries and interfaces with other closed TFTs is beyond the scope of this note.
}

{\bf Q3:} {\it Given a fusion category, is there a (1+1)$d$ TFT whose TDLs (or a subset thereof) realize the said category?}

Thorngren and Wang \cite{Thorngren:2019iar} has proven that $\cC$-symmetric defect TFTs are in bijection with $\cC$-module categories, and since the regular module category always exists, {\bf Q3} has been affirmatively answered.  However, their proof utilizes the (1+1)$d$ Turaev-Viro state-sum \cite{Turaev:1992hq} or Levin-Wen string-net \cite{Levin:2004mi} construction, and it is generally unclear how the axiomatic TFT data can be extracted.  We are thus led to the next question:

{\bf Q4:} {\it Given a fusion category, can one construct the axiomatic data of a (1+1)$d$ TFT whose TDLs (or a subset thereof) realize the said category?}

This question has been answered for group-like categories by Wang, Wen, and Witten \cite{Wang:2017loc} and by Tachikawa \cite{Tachikawa:2017gyf}, and for categories with fiber functors (the resulting TFT has a unique vacuum) by Thorngren and Wang \cite{Thorngren:2019iar}.  In \cite{Chang:2018iay}, it was shown that for a variety of CFTs, the TFT data can be solved solely from the input of the fusion categorical data, by bootstrapping the consistency conditions.  For general categories, a construction of the bulk Frobenius algebra was given by Komargodski, Ohmori, Roumpedakis, and Seifnashri \cite{Komargodski:2020mxz}, but the full defect TFT data remains unsolved.\footnote{After publication of the first version of this note, Kantaro Ohmori and Sahand Seifnashri suggested to the authors that a construction of the full defect TFT data may be possible through a generalization of \cite{Komargodski:2020mxz}.
}

The preceding questions are ultimately connected.  A CFT realizing a certain fusion category is connected to a TFT realizing the same category under TDL-preserving renormalization group (RG) flows, and this principle strongly constrains the infrared fate of CFTs.  In the space of TDL-preserving RG flows and without fine-tuning, they must either flow to gapped phases described by TFTs, or to ``dead-end'' CFTs \cite{Nakayama:2015bwa}, which correspond to gapless phases protected by fusion categories \cite{Thorngren:2019iar}.\footnote{Such phases  generalize the notion of (group-like) symmetry-protected gapless phases \cite{Chen:2011bcp} and perfect metals \cite{Plamadeala:2014roa}.
}

As desirable as fully universal answers to the preceding questions are, a more pragmatic approach may be to first examine fusion categories for which the answers are not known.  This note makes a modest offering in this approach of pursuing exotica in the quest for their eventual conformity: the construction of a TFT realizing the Haagerup ${\cal H}_3$ fusion category with fully explicit axiomatic TFT data.  The construction is of bootstrap nature, by solving the full cutting and sewing consistency conditions. A prerequisite in this approach is the explicit knowledge of the $F$-symbols.  They were implicit in the work of Grossman and Snyder \cite{Grossman_2012} (using a generalization of the approach by Izumi \cite{izumi2001structure}), and also explicitly obtained by Titsworth \cite{Titsworth}, Osborne, Stiegemann, and Wolf \cite{osborne2019fsymbols}.  In \cite{Huang:2020lox}, the present authors recast the $F$-symbols in a gauge that manifests the transparent property, which greatly simplify our present computational endeavor.

The remaining sections are organized into steps of the construction and discussions of further ramifications.  Section~\ref{Sec:GeneralTFT} reviews the generalities of topological field theory extended by defects, and formulates the defining data and consistency conditions.  Section~\ref{Sec:Modular} introduces the Haagerup fusion ring with six simple objects/TDLs, studies its representation theory, and constrains the vacuum degeneracy using modular invariance.  Section~\ref{Sec:Z3} studies the relations among dynamical data implied by transparency and $\bZ_3$ symmetry.  Section~\ref{Sec:Consistency} delineates the constraints of associativity and torus one-point modular invariance.  Section~\ref{Sec:TFT} solves the constraints to construct a topological field theory with Haagerup symmetry.  Section~\ref{Sec:NIM} examines the expectation that the boundary conditions furnish a non-negative integer matrix representation (NIM-rep) of the fusion ring.  Section~\ref{Sec:Gauging} discusses the relations among topological field theories by gauging algebra objects.  Section~\ref{Sec:Remarks} ends with some prospective questions.  Appendix~\ref{App:CrossingKernel} contains the $F$-symbols for the Haagerup $\cH_3$ fusion category.  Appendix~\ref{App:Defect} analyzes the general crossing symmetry of defect operators.

\section{Topological field theory extended by defects}
\label{Sec:GeneralTFT}

This section introduces the axiomatic data of a topological field theory (TFT) extended by defects, and the consistency conditions they must satisfy.

\subsection{Fusion category of topological defect lines}

The nontrivial splitting and joining relations of a finite set of topological defect lines (TDLs) are captured by a fusion category.  A classic introduction to fusion categories can be found in \cite{Etingof:aa,etingof2016tensor}, and expositions in the physics context can be found in \cite{Bhardwaj:2017xup,Chang:2018iay}.  Here we follow the latter and present a lightening review of the key properties of TDLs.

Topological defect lines are (generally oriented) defect lines whose isotopic transformations leave physical observables invariant.  We restrict ourselves to considering sets of TDLs with finitely many {\it simple} TDLs $\{ \cL_i \}$; the others, the {\it non-simple} TDLs, are direct sums of the simple ones.\footnote{See \cite{Chang:2020imq} for progress in incorporating ``non-compact'' topological defect lines.
}
Among the simple TDLs there is a trivial TDL $\cI$ representing nothingness.  Furthermore, every TDL $\cL$ has an orientation reversal $\ocL$, as depicted by the equivalence
\ie
\begin{gathered}
\begin{tikzpicture}[scale=.5]
\draw [line,->-=.55] (0,-1.5) -- (0,0) node [left] {$\cL$} -- (0,1.5);
\end{tikzpicture}
\end{gathered} 
\quad = \quad
\begin{gathered}
\begin{tikzpicture}[scale=.5]
\draw [line,-<-=.55] (0,-1.5) -- (0,0) node [right] {$\ocL$} -- (0,1.5);
\end{tikzpicture}
\end{gathered} 
\, .
\fe
Whenever a TDL is isomorphic to its own orientation reversal, $\cL = \ocL$, we omit the arrows on the lines.\footnote{The orientation cannot be completely ignored if the TDL has an orientation-reversal anomaly (nontrivial Frobenius-Schur indicator) \cite{Chang:2018iay}.  This subtlety does not arise for the Haagerup and is therefore neglected.
}

A general configuration of TDLs involves junctions built out of trivalent vertices.  The allowed trivalent vertices are specified by the fusion ring
\ie
\cL_i \cL_j = N_{ij}^k \cL_k \, ,
\fe
where $N_{ij}^k \in \bZ_{\ge0}$ are the fusion coefficients.  To simplify the discussion, it is assumed that (1) the fusion coefficients (dimensions of junction vector spaces) are zero or one, and (2) the trivalent vertices are cyclic-permutation invariant.\footnote{Both assumptions are satisfied by the transparent Haagerup $\cH_3$ fusion category.  The reader is referred to \cite{Chang:2018iay} for a general discussion without these assumptions.
}
In conformity with \cite{Chang:2018iay,Huang:2020lox}, we adopt the counter-clockwise convention for trivalent vertices, such that
\ie
\begin{gathered}
\begin{tikzpicture}[scale=1]
\draw [line,->-=1] (0,0) 
-- (0,1) node [above] {$\cL_3$};
\draw [line,->-=1] (0,0) -- (-.87,-.5) node [below left] {$\cL_1$};
\draw [line,->-=1] (0,0) -- (.87,-.5) node [below right] {$\cL_2$};
\end{tikzpicture}
\end{gathered}
\fe
is allowed when $\cI \in \cL_1 \cL_2 \cL_3$.  To completely specify a trivalent vertex, a junction vector must be chosen from the junction vector space $V_{\cL_1, \cL_2, \cL_3}$.\footnote{In the path integral language, a junction vector specifies the boundary conditions of quantum fields at a trivalent vertex.
}
The collection of choices for all trivalent vertices formed by all simple TDLs constitutes a gauge.  

The fusion product of a simple TDL $\cL$ with its orientation reversal contains the trivial TDL, 
\ie
\cL \, \ocL = \cI + \dotsb \, ,
\fe
because clearly
\ie
\begin{gathered}
\begin{tikzpicture}[scale=1]
\draw [line,dashed] (0,0) 
-- (0,1) node [above] {$\cI$};
\draw [line,->-=1] (0,0) -- (-.87,-.5) node [below left] {$\cL$};
\draw [line,->-=1] (0,0) -- (.87,-.5) node [below right] {$\ocL$};
\end{tikzpicture}
\end{gathered}
~~=~~
\begin{gathered}
\begin{tikzpicture}[scale=1]
\draw (0,1.25) node [above] {~};
\draw [line,->-=1] (0,0) -- (-.87,-.5) node [below left] {$\cL$};
\draw [line] (0,0) -- (.87,-.5);
\end{tikzpicture}
\end{gathered}
\fe
is allowed.  Another important notion is {\it invertibility}.  A TDL $\cL$ is invertible if $\cL \, \ocL = \cI$, and non-invertible otherwise.  Invertible TDLs are equivalent to background gauge bundles for finite symmetry groups \cite{Gaiotto:2014kfa,Bhardwaj:2017xup}.

The splitting and joining of TDLs can be decomposed into basic $F$-moves that are characterized by the $F$-symbols.  In a given gauge, the $F$-symbols are $\bC^\times$-numbers, and an $F$-move is the equivalence between the two configurations
\ie
\label{F}
\begin{gathered}
\begin{tikzpicture}[scale=1]
\draw [line,-<-=.56] (-1,0) 
-- (-.5,0) node [above] {$\cL_5$} -- (0,0);
\draw [line,->-=1] (-1,0) -- (-1.5,.87) node [above left=-3pt] {$\cL_1$};
\draw [line,->-=1] (-1,0) -- (-1.5,-.87) node [below left=-3pt] {$\cL_2$};
\draw [line,->-=1] (0,0) -- (.5,-.87) node [below right=-3pt] {$\cL_3$};
\draw [line,->-=1] (0,0) 
-- (.5,.87) node [above right=-3pt] {$\cL_4$};
\end{tikzpicture}
\end{gathered}
~~=~~
\sum_{\cL_6}
(F^{\cL_1, \cL_2, \cL_3}_{\ocL_4})_{\cL_5, \cL_6}
\begin{gathered}
\begin{tikzpicture}[scale=1]
\draw [line,-<-=.56] (0,-1) 
-- (0,-.5) node [right] {$\cL_6$} -- (0,0);
\draw [line,->-=1] (0,-1) -- (.87,-1.5) node [below right=-3pt] {$\cL_3$};
\draw [line,->-=1] (0,-1) -- (-.87,-1.5) node [below left=-3pt] {$\cL_2$};
\draw [line,->-=1] (0,0) -- (-.87,.5) node [above left=-3pt] {$\cL_1$};
\draw [line,->-=1] (0,0) 
-- (.87,.5) node [above right=-3pt] {$\cL_4$};
\end{tikzpicture}
\end{gathered}
\, .
\fe
The $F$-symbols must satisfy the pentagon identity, which can only have finitely many solutions (up to gauge equivalence) for a given fusion ring due to Ocneanu rigidity \cite{wassermann2000quantum,Etingof:aa}.

\subsection{Local operators and commutative Frobenius algebra}
\label{Sec:Frobenius}

Topological defect lines act on local operators by circling and shrinking.  In conformity with \cite{Chang:2018iay,Huang:2020lox}, we adopt the clockwise convention for action on local operators,
\ie
\label{TDLAction}
\begin{gathered}
\begin{tikzpicture}[scale = .75]
\draw [line,-<-=.25] (0,0) \dt{below}{0}{$\cO$} circle (1);
\draw (-1,0) -- (-1,0) node[left] {$\cL$};
\end{tikzpicture}
\end{gathered}
~=~
\wh\cL(\cO) \, .
\fe
For instance, if $\cO_q$ is a local operator with $\bZ_3$-charge $q$, and if $\A$ is the TDL corresponding to the generator of $\bZ_3$, then
\ie
\begin{gathered}
\begin{tikzpicture}[scale = .75]
\draw [line,-<-=.25] (0,0) \dt{below}{0}{$\cO_q$} circle (1);
\draw (-1,0) -- (-1,0) node[left] {$\A$};
\end{tikzpicture}
\end{gathered}
~=~
\omega^q \, \cO_q \, .
\fe

The data of local operators is captured by a commutative Frobenius algebra \cite{Abrams:1996ty,kock2004frobenius}.  Commutativity guarantees that a projector basis exists:
\ie
\{ \pi_a, \, a = 1, \dotsc, n_{\rm V} \mid \pi_a \pi_b = \D_{ab} \pi_a \} \, ,
\fe
where $n_{\rm V}$ denotes the number of vacua.  In this basis, the nontrivial data is captured in the overlap of the projectors with the identity, {\it i.e.} the one-point functions $\la \pi_a \ra$.  Most of this note does not work in the projector basis, because for us it is more convenient to work in a basis that simplifies the TDL actions as much as possible.  However, the projector basis will figure in the discussion of boundary states in Section~\ref{Sec:NIM}.

\subsection{Defect operators, defect operator algebra, and lassos}
\label{Sec:DefectOp}

Associated to every topological defect line $\cL$ is a defect Hilbert space $\cH_\cL$, which contains states quantized on a spatial circle with twisted periodic boundary conditions. Via the state-operator map, 
\ie
\begin{tikzpicture}[scale=.5]
\draw [bg] (0,0) ++ (180:2 and 1) arc (180:360:2 and 1);
\draw [bg,dashed] (0,0) ++ (0:2 and 1) arc (0:180:2 and 1);
\draw [bg] (0,4) ellipse (2 and 1);
\draw [bg] (-2,0) -- (-2,4);
\draw [bg] (2,0) -- (2,4);
\draw [line,->-=.77] (0,-1) node[below] {$|\cO\rangle$} -- (0,1) -- node[left] {$\cal L$} (0,3);
\node at (4.5,1.5) {$\mapsto$};
\draw [line,->-=.52] (8,0) \dt{below}{0}{$\cO(x)$}
-- node[left] {$\cal L$} (8,4);
\node at (11,-1) {};
\end{tikzpicture}
\fe
$\cH_\cL$ is also the Hilbert space of point-like {\it defect operators} on which $\cL$ can end.  Defect Hilbert spaces are equipped with a norm
\ie
{\cal H}_{\cL} \otimes {\cal H}_{\ocL} \to \bC \, ,
\fe
which defines a hermitian structure.  The hermitian conjugate of $\cO$ will be denoted by $\ocO$.

The spectral data of a topological field theory extended by defects consists of the set of local operators, their representations under the fusion ring, and the set of defect operators. The dynamical data consists of the operator product
\ie
\cO_1 \otimes \cO_2 \in \cH_{\ocL_1} \otimes \cH_{\ocL_2}
~~\mapsto~~
\begin{gathered}
\begin{tikzpicture}[scale = 1]
\draw [line,-<-=.6] (-1,0) node[left] {$\cL_3$} -- (0,0);
\draw [line,->-=.6] (0,0) -- (.25,-.43) node[below left=-3pt] {$\cL_1$} -- (.5,-.87) \dt{below right}{-3}{$\cO_1$};
\draw [line,->-=.6] (0,0) -- (.25,.43) node[above left=-2pt] {$\cL_2$} -- (.5,.87) \dt{above right}{-3}{$\cO_2$};
\end{tikzpicture}
\end{gathered}
~~ \in \cH_{\cL_3}
\fe
and the lasso action
\ie
\cO_4 \in \cH_{\cL_4} ~~\mapsto~~
\lassoo{1}{$\cL_1$}{$\cL_2$}{$\cL_3$}{$\cO_4$}{}{$\cL_4$} ~~ \in \cH_{\cL_1} \, .
\fe
When $\cL_1 = \cL_4 = \cI$ and $\cL_2 = \ocL_3$, the above diagram becomes \eqref{TDLAction}, and the lasso action reduces to the TDL action $\wh\cL_2$ on local operators that maps $\cH$ to $\cH$.  The lasso action is a generalization that maps a defect Hilbert space $\cH_4$ to another defect Hilbert space $\cH_1$.  In the following, for TDLs ending on defect operators, the labeling of the former will be suppressed as it is implied by that of the latter.

The closest analog of charge conservation for a non-invertible TDL $\cL$ is to circle a pair of local operators by $\cL$, and impose the commutativity of (1) taking the local operator product and (2) performing an $F$-move and studying the defect operator product, as illustrated below: 
\ie
\label{ChargeConservation}
\hspace{-.2in}
\begin{gathered}
\begin{tikzpicture}[every text node part/.style={align=center}]
\coordinate (TL)  at (-4,4);
\draw [->,thick] (TL)++(3,0) -- +(4,0);
\draw [->,thick] (TL)+(0, -1.5)--+(0,-3);
\draw [->,thick] (6,2.5)--(6,1);
\draw (1.25,.05)--(2.5,.05) node[above=5pt]{``~~~~charge~~~~''\\conservation} --(3.75,.05);
\draw (1.25,-.05)--(2.5,-.05) --(3.75,-.05);
\draw [line] (TL) ellipse (2 and 1);
\draw [line] (TL)+(-0.5,0) \dt{below}{0}{$\cO_1$};
\draw [line] (TL)+(0.5,0) \dt{below}{0}{$\cO_2$};
\draw [line] (6,4) ellipse (2 and 1);
\draw [line] (6,4) \dt{below}{0}{$\cO_1\times\cO_2$};
\draw [line] (-3.1,0) \dt{below}{0}{$\cO_1$} circle (.7);
\draw [line] (-2.4, 0) -- (-2,0) node[above] {$\cL'$}-- (-1.6, 0);
\draw [line] (-0.9,0) \dt{below}{0}{$\cO_2$}circle (.7);
\node at (-5.7,-0.2) {$\displaystyle\sum_{\cL'}(F_{\bar\cL}^{\cL, \bar\cL, \bar\cL})_{I,\cL'}$};
\node at (-2,4.7) {$\cL$};
\node at (8,4.7) {$\cL$};
\node at (-4.1,0) {$\cL$};
\node at (0.1,0) {$\cL$};
\node at (6,-0.2) {$\displaystyle\sum_{\cO\in \cO_1\times \cO_2} \wh{\cL}(\cO)$};
\end{tikzpicture}
\end{gathered}
\fe

By the use of the norm, the operator product is equivalently encoded in the three-point coefficients
\ie
c(\cO_1, \cO_2, \cO_3) ~~ = ~~
\tri{.5}{1}{$\cO_1$}{$\cO_2$}{$\cO_3$} ~~ \in \bC \, ,
\fe 
and the lasso action is encoded in the lasso coefficients
\ie
\lasso{.5}{$\cO_1$}{$\cL_2$}{$\cL_3$}{$\cO_4$} ~~ \in \bC \, .
\fe
In the above, vacuum expectation values are implicitly taken.  The three-point coefficients are invariant under cyclic permutations
\ie
\label{InvCyclic}
c(\cO_1, \cO_2, \cO_3) = c(\cO_2, \cO_3, \cO_1) = c(\cO_3, \cO_1, \cO_2) \, ,
\fe
and complex conjugate under reflections
\ie
\label{ConjReflect}
c(\cO_1, \cO_2, \cO_3) = c(\ocO_1, \ocO_3, \ocO_2)^* \, .
\fe
The lasso coefficients also enjoy the symmetries
\ie
\lasso{.5}{$\cO_1$}{$\cL_2$}{$\cL_3$}{$\cO_4$} 
&= \lasso{.5}{$\cO_4$}{$\ocL_3$}{$\ocL_2$}{$\cO_1$}
=
\left( \lasso{.5}{$\ocO_1$}{$\ocL_2$}{$\ocL_3$}{$\ocO_4$} \right)^* \, .
\fe

\subsection{General observables, crossing symmetry and modular invariance}
\label{Sec:Crossing}

A general observable in a topological field theory extended by defects is the vacuum expectation value of a graph---a configuration of topological defect lines with junctions and endpoints---on a Riemann surface.\footnote{Each observable can be interpreted as a transition amplitude over some time function, with nontrivial topology changes and defect dressing.  See \cite{Bhardwaj:2017xup} for an exposition from this perspective.
}
On the sphere, any graph can be expanded into a sum of local operators, and taking the vacuum expectation value amounts to computing the overlap with the identity.  The basic building blocks for this computation are the three-point and lasso coefficients introduced earlier, and the computation also involves basic manipulations of TDLs such as $F$-moves.  Observables on general Riemann surfaces can be reduced to those on the sphere by a pair-of-pants decomposition.  The equivalence of the various ways of building the same observable on a general Riemann surface is guaranteed by the four-point crossing symmetry and torus one-point modular invariance \cite{Chang:2018iay}, generalizing the situation without defects argued by Sonoda \cite{Sonoda:1988mf,Sonoda:1988fq} and by Moore and Seiberg \cite{Moore:1988qv,Moore:1989vd}.  In the following, all local and defect operators are taken to be canonically normalized under the hermitian structure,
\ie
\langle \defect{.5}{1}{$\cL$}{$\cO$}{$\ocO$} \rangle = 1 \, .
\fe

On the sphere, the four-point correlator of local and defect operators $\cO_i \in {\cal H}_{\cL_i}$ bridged by an internal $\cL \in \cL_1 \cL_2 \cap \ocL_4 \ocL_3$ can be decomposed into three-point coefficients by cutting across $\cL$ (with the cut shown by the dotted lines),
\ie
\begin{gathered}
\begin{tikzpicture}[scale = 1]
\draw [line,-<-=.6] (-1,0) -- (-.5,0) node[above] {$\cL$} -- (0,0);
\draw [line,->-=.6] (-1,0) -- (-1.5,.87) \dt{above left}{-3}{$\cO_1$};
\draw [line,->-=.6] (-1,0) -- (-1.5,-.87) \dt{below left}{-3}{$\cO_2$};
\draw [line,->-=.6] (0,0) -- (.5,-.87) \dt{below right}{-3}{$\cO_3$};
\draw [line,->-=.6] (0,0) -- (.5,.87) \dt{above right}{-3}{$\cO_4$};
\draw [line,dotted] (-.2,-1.45) -- (-.2,1.45);
\end{tikzpicture}
\end{gathered}
~~ = \sum_{\cO \in \cH_\cL} c(\cO_1, \cO_2, \cO) \, c(\cO_3, \cO_4, \ocO) \, .
\fe
Under an $F$-move,
\ie
\begin{gathered}
\begin{tikzpicture}[scale = 1]
\draw [line,-<-=.6] (-1,0) -- (-.5,0) node[above] {$\cL$} -- (0,0);
\draw [line,->-=.6] (-1,0) -- (-1.5,.87) \dt{above left}{-3}{$\cO_1$};
\draw [line,->-=.6] (-1,0) -- (-1.5,-.87) \dt{below left}{-3}{$\cO_2$};
\draw [line,->-=.6] (0,0) -- (.5,-.87) \dt{below right}{-3}{$\cO_3$};
\draw [line,->-=.6] (0,0) -- (.5,.87) \dt{above right}{-3}{$\cO_4$};
\draw [line,dotted] (-.2,-1.45) -- (-.2,1.45);
\end{tikzpicture}
\end{gathered}
~~ = \sum_{\cL'} ~~
\begin{gathered}
\begin{tikzpicture}[scale = 1]
\draw [line,-<-=.6] (0,-1) -- (0,-.5) node[right] {${\cL}'$} -- (0,0);
\draw [line,->-=.6] (0,-1) -- (.87,-1.5) \dt{below right}{-3}{$\cO_3$};
\draw [line,->-=.6] (0,-1) -- (-.87,-1.5) \dt{below left}{-3}{$\cO_2$};
\draw [line,->-=.6] (0,0) -- (-.87,.5) \dt{above left}{-3}{$\cO_1$};
\draw [line,->-=.6] (0,0) -- (.87,.5) \dt{above right}{-3}{$\cO_4$};
\draw [line,dotted] (-1.45,-.2) -- (1.45,-.2);
\end{tikzpicture}
\end{gathered} 
~~
(F^{\cL_1, \cL_2, \cL_3}_{\ocL_4})_{\cL, \cL'}
\, ,
\fe
where each graph appearing on the right can be decomposed into three-point coefficients by cutting across $\cL'$.  Crossing symmetry is the equivalence of
\ie
\hspace{-.1in} \sum_{\cO \in \cH_\cL} c(\cO_1, \cO_2, \cO) \, c(\cO_3, \cO_4, \ocO) = \sum_{\cL'} (F^{\cL_1, \cL_2, \cL_3}_{\ocL_4})_{\cL, \cL'} \sum_{\cO' \in \cH_{\cL'}} c(\cO_2, \cO_3, \cO') \, c(\cO_4, \cO_1, \ocO') \, .
\fe

The modular invariance of the torus one-point function begins with performing $F$-moves on the configuration
\ie
\begin{gathered}
\begin{tikzpicture}[scale = 1.5]
\draw [bg] (0,0) -- (2,0) -- (2,2) -- (0,2) -- (0,0);
\draw (1,0) node[below] {$\cL_1$};
\draw (2,1) node[right] {$\cL_2$};
\draw (1,2) node[above] {$\cL_1$};
\draw (0,1) node[left] {$\cL_2$};
\draw [line,->-=.3,->-=.8] (2,0) ++(180:1) arc (180:90:1);
\draw [line,->-=.3,->-=.8] (0,2) ++(-90:1) arc (-90:0:1);
\draw [line,->-=.3,-<-=.8] (1.29,0.71) -- (1.15,0.85) node[above right=-2pt] {$\cL_3$} -- (0.85,1.15) node[above right=-2pt] {$\cL_4$} -- (0.71,1.29);
\draw [line,->-=.6] (1,1) -- (0.8,0.8) 
-- (0.6,0.6) \dt{below left}{-5}{$\cO$};
\end{tikzpicture}
\end{gathered}
\quad &= \quad
\sum_{\cL' \in \cL_\cO \ocL_1}
(F^{\ocL_4, \cL_\cO, \ocL_1}_{\ocL_2})_{\cL_3, \cL'}
\quad
\begin{gathered}
\begin{tikzpicture}[scale = 1.5]
\draw [bg] (0,0) -- (2,0) -- (2,2) -- (0,2) -- (0,0);
\draw (1,0) node[below] {$\cL_1$};
\draw (2,1) node[right] {$\cL_2$};
\draw (1,2) node[above] {$\cL_1$};
\draw (0,1) node[left] {$\cL_2$};
\draw [line,->-=.5] (2,0) ++(180:1) arc (180:157.5:1);
\draw [line,->-=.5] (2,0) ++(157.5:1) arc (157.5:135:1) node[below] {$\cL'$};
\draw [line,->-=.5] (2,0) ++(135:1) arc (135:90:1);
\draw [line,->-=.3,->-=.8] (0,2) ++(-90:1) arc (-90:0:1);
\draw [line,-<-=.5] (1.29,0.71) --  node[above right=-2pt] {$\cL_4$} (0.71,1.29);
\draw [line,->-=.6] (1.08,0.38) -- (0.6,0.6) \dt{below left}{-5}{$\cO$};
\draw [line,dotted] (-.5,.67)--(2.5,.67);
\draw [line,dotted] (-.5,1.75)--(2.5,1.75);
\end{tikzpicture}
\end{gathered}
\\
& 
= \quad
\sum_{\cL' \in \ocL_2 \cL_\cO}
(F^{\cL_1, \ocL_2, \cL_\cO}_{\cL_3})_{\ocL_4, \cL'}
\quad
\begin{gathered}
\begin{tikzpicture}[scale = 1.5]
\draw [bg] (0,0) -- (2,0) -- (2,2) -- (0,2) -- (0,0);
\draw (1,0) node[below] {$\cL_1$};
\draw (2,1) node[right] {$\cL_2$};
\draw (1,2) node[above] {$\cL_1$};
\draw (0,1) node[left] {$\cL_2$};
\draw [line,->-=.3,->-=.8] (2,0) ++(180:1) arc (180:90:1);
\draw [line,->-=.5] (0,2) ++(-90:1) arc (-90:-67.5:1);
\draw [line,->-=.5] (0,2) ++(-67.5:1) arc (-67.5:-45:1)node[left] {$\cL'$};
\draw [line,->-=.5] (0,2) ++(-45:1) arc (-45:0:1);
\draw [line,->-=.5] (1.29,0.71) --  node[above right=-2pt] {$\cL_3$} (0.71,1.29);
\draw [line,->-=.6] (0.38,1.08) -- (0.6,0.6) \dt{below left}{-5}{$\cO$};
\draw [line,dotted] (.67,-.5)--(.67,2.5);
\draw [line,dotted] (1.75,-.5)--(1.75,2.5);
\end{tikzpicture}
\end{gathered}
\fe
and demanding the equivalence of the two cuts shown by the dotted lines:
\ie
\label{ModularInvariance}
& \sum_{\cL' \in \cL_\cO \ocL_1} \sum_{\cO_1 \in {\cal H}_{\cL_1}} \sum_{\cO' \in {\cal H}_{\cL'}} ~ (F^{\ocL_4, \cL_\cO, \ocL_1}_{\ocL_2})_{\cL_3, \cL'} ~ c(\cO, \ocO_1, \ocO') ~ \lasso{.5}{$\cO_1$}{$\cL_2$}{$\ocL_4$}{$\cO'$}
\\
= & \sum_{\cL' \in \ocL_2 \cL_\cO} \sum_{\cO_2 \in {\cal H}_{\cL_2}} \sum_{\cO' \in {\cal H}_{\cL'}} ~ (F^{\cL_1, \ocL_2, \cL_\cO}_{\cL_3})_{\ocL_4, \cL'} ~ c(\cO, \ocO_2, \ocO') ~ \lasso{.5}{$\cO_2$}{$\cL_1$}{$\ocL_3$}{$\cO'$} \, .
\fe

\section{Spectral constraints by Haagerup symmetry}
\label{Sec:Modular}

This section studies the modular constraints on the spectral data---the set of local operators, their representations under the fusion ring, and the set of defect operators---when the theory is known to contain topological defect lines (TDLs) realizing the Haagerup $\cH_3$ fusion category.

\subsection{The Haagerup fusion ring with six simple objects}
\label{Sec:Fusion}

The Haagerup $\cH_3$ fusion category was constructed by Grossman and Snyder \cite{Grossman_2012} as a variant (Grothendieck equivalent) of the $\cH_2$ fusion category that directly came from the Haagerup subfactor \cite{asaeda1999exotic,haagerup1994principal}.  There are six simple objects/TDLs, which we denote by
\ie
\cI, \quad \A, \quad \A^2, \quad \rho, \quad \A\rho, \quad \A^2\rho \, .
\fe
The fusion ring is fully specified by the relations
\ie
\A^3 = 1 \, , \quad \A \rho = \rho \, \A^2 \, , \quad \rho^2 = \cI + \cZ \, , \quad \cZ \equiv \sum_{i=0}^2 \A^i \rho \, .
\fe
For shorthand,
\ie
\rho_i \equiv \A^i \rho \, .
\fe
In the rest of this note, we use unoriented solid lines to denote the non-invertible self-dual simple TDLs $\rho_i$, and oriented dashed lines to denote the invertible ones:
\ie
\begin{gathered}
\begin{tikzpicture}[scale = 1]
\draw [line,dashed,->-=.5] (0,-1) -- (0,1);
\end{tikzpicture}
\end{gathered}
\quad = \quad
\begin{gathered}
\begin{tikzpicture}[scale = 1]
\draw [line,->-=.5] (0,-1) -- (0,0)node[right=2pt] {$\alpha$} -- (0,1) ;
\end{tikzpicture}
\end{gathered}
\, ,
\qquad\qquad
\begin{gathered}
\begin{tikzpicture}[scale = 1]
\draw [line,dashed,-<-=.5] (0,-1) -- (0,1);
\end{tikzpicture}
\end{gathered}
\quad = \quad
\begin{gathered}
\begin{tikzpicture}[scale = 1]
\draw [line,->-=.5] (0,-1) -- (0,0)node[right=2pt] {$\bar\alpha$} -- (0,1) ;
\end{tikzpicture}
\end{gathered}
\, ,
\qquad\qquad
\begin{gathered}
\begin{tikzpicture}[scale = 1]
\draw [line] (0,-1) -- (0,0)node[right=2pt] {$\rho_i$} -- (0,1) ;
\end{tikzpicture}
\end{gathered} \, .
\fe

There are two gauge-inequivalent unitary fusion categories realizing the above fusion ring, denoted $\cH_2$ and $\cH_3$ by Grossman and Snyder \cite{Grossman_2012}.  Whereas the Haagerup $\cH_2$ fusion category descended directly from the Haagerup subfactor \cite{asaeda1999exotic,haagerup1994principal}, the Haagerup $\cH_3$ fusion category was constructed by Grossman and Snyder \cite{Grossman_2012} based on $\cH_2$.  It turns out to be easier to work with $\cH_3$, but the analysis in this section applies to both $\cH_2$ and $\cH_3$.  The $F$-symbols for $\cH_3$ were implicit in the work of Grossman and Snyder \cite{Grossman_2012} (using a generalization of the approach by Izumi \cite{izumi2001structure} for $\cH_2$), and also explicitly obtained by Titsworth \cite{Titsworth}, Osborne, Stiegemann, and Wolf \cite{osborne2019fsymbols}.  In \cite{Huang:2020lox}, the present authors recast the $F$-symbols in a gauge that manifests the transparent property, a notion we introduce in Section~\ref{Sec:Z3}.  The transparent $F$-symbols are given in Appendix~\ref{App:CrossingKernel}.

\subsection{Action on local operators and representation theory}

To describe how topological defect lines forming the Haagerup $\cH_3$ fusion category act on local operators, we should first study the complex representation theory of its fusion ring.  Since the fusion ring is non-commutative, the action of TDLs cannot be simultaneously diagonalized. We work in a basis in which the action of $\bZ_3$ is diagonal.

\begin{itemize}
\item
For a state $|\phi\ra$ neutral under $\bZ_3$,
\ie
\rho |\phi\ra = \A\rho |\phi\ra = \A^2\rho |\phi\ra \, , \quad \cZ |\phi\ra = 3 \rho |\phi\ra \, ,
\fe
hence there are two one-dimensional representations,
\ie
\rho |\phi\ra = {3 \pm \sqrt{13} \over 2} |\phi\ra \, .
\fe

\item
For a state $|\phi\ra$ with unit $\bZ_3$-charge,
\ie
\A |\phi\ra = \omega |\phi\ra \, ,
\quad
\A\rho |\phi\ra = \rho\A^2 |\phi\ra = \omega^2 \rho |\phi\ra \, ,
\quad
\A^2\rho |\phi\ra = \rho\A |\phi\ra = \omega \rho |\phi\ra \, .
\fe
It follows that $\cZ |\phi\ra = 0$, and hence
\ie
\rho^2 |\phi\ra = |\phi\ra \, .
\fe
If $\rho |\phi\ra$ and $|\phi\ra$ were equal up to a phase, then there would be two possible one-dimensional representations with
\ie
\rho |\phi\ra = \pm |\phi\ra \, ,
\fe
which is in conflict with $\A \rho = \rho \A^2$.  Hence $\rho |\phi\ra$ and $|\phi\ra$ must be independent, and the representation is two-dimensional.  In the $(|\phi\ra, \, \rho |\phi\ra)$ basis,
\ie
\label{Rep2}
\A = \begin{pmatrix}
\omega & 0
\\
0 & \omega^2
\end{pmatrix} \, , \quad 
\rho = \begin{pmatrix}
0 & 1
\\
1 & 0
\end{pmatrix} \, .
\fe
\end{itemize}

The above classification of irreducible representations is summarized in Table~\ref{Tab:Rep}.  In a reflection-positive quantum field theory, the identity operator transforms in a one-dimensional representation with positive charges.  Here, under the reflection-positive assumption, the identity operator must transform in the $+$ representation.

\begin{table}[H]
\centering
\begin{tabular}{c|cc}
$\bf r$ & $\A$ & $\rho$
\\\hline
$+$ & 1 & ${3+\sqrt{13} \over 2}$
\\
$-$ & 1 & ${3-\sqrt{13} \over 2}$
\\
{\bf 2} & 
$\begin{pmatrix} \omega & 0 \\ 0 & \omega^2 \end{pmatrix}$ & 
$\begin{pmatrix} 0 & 1 \\ 1 & 0 \end{pmatrix}$
\end{tabular}
\caption{Irreducible representations of the Haagerup fusion ring with six simple objects/TDLs.}
\label{Tab:Rep}
\end{table}

\subsection{Modular invariance and vacuum degeneracy}

Let $n_{\rm V}$ denote the number of vacua (local operators), and $n_\pm$ and $n_{\bf 2}$ be their multiplicities of representations (in the notation of Table~\ref{Tab:Rep}).  Clearly, $n_{\rm V} = n_+ + n_- + 2 n_{\bf 2}$.

Consider the modular invariance of the torus partition function with the non-invertible TDL $\rho$ wrapped around a one-cycle
\ie
\label{RhoT1Pt}
\begin{gathered}
\begin{tikzpicture}[scale = 1.5]
\draw [bg] (0,0) -- (2,0) -- (2,2) -- (0,2) -- (0,0);
\draw [line] (0,1) -- (1,1) node[above] {$\rho$} -- (2,1);
\draw [line,dotted] (-.5,.25)--(2.5,.25);
\draw [line,dotted] (.25,-.5)--(.25,2.5);
\end{tikzpicture}
\end{gathered}
\fe
The horizontal cut computes the trace over the action of $\wh\rho$ in the Hilbert space $\cH$ of local operators, and the vertical cut simply counts the dimensionality of the defect Hilbert space $\cH_\rho$.  Modular invariance requires
\ie
\label{XiModularInvariance}
{\rm Tr}_{\cal H} \, \wh{\rho} 
= {\rm Tr}_{{\cal H}_{\rho}} 1 \in \bZ_{\ge0} \, .
\fe
Given the representation content of the Haagerup fusion ring, summarized in Table~\ref{Tab:Rep}, we immediately conclude that $n_+ = n_-$, and the number of vacua must be even.  Let us write
\ie
n_{\bf 1} \equiv n_+ = n_-
\fe
to denote the multiplicity of each one-dimensional representation.  Using the ${\rm U}(n_{\bf 2})$ freedom, we can choose a basis of local operators to represent $\wh\rho$ in block diagonal form
\ie
\wh\rho &= \bigoplus_{p=1}^{n_+} \begin{pmatrix} \displaystyle {3 + \sqrt{13} \over 2} \end{pmatrix} \oplus \bigoplus_{q=1}^{n_-} \begin{pmatrix} \displaystyle {3 - \sqrt{13} \over 2} \end{pmatrix} \oplus
\begin{pmatrix}
0 & 1
\\
1 & 0
\end{pmatrix} 
\oplus \dotsb \oplus
\begin{pmatrix}
0 & 1
\\
1 & 0
\end{pmatrix} \, .
\fe
Modular invariance \eqref{XiModularInvariance} also implies that the defect Hilbert space $\cH_\rho$ is $3 n_{\bf 1}$-dimensional, {\it i.e.} the TDL $\rho$ can end on 
\ie
n_\rho = 3 n_{\bf 1}
\fe
independent defect operators.  And similarly for each of the other $\rho_i$.

\begin{table}
\centering
\begin{tabular}{|c|c|c||c|c||c|}
\hline
$n_{\rm V}$ & $n_{\bf 1} = n_+ = n_-$ & $n_{\bf 2}$ & $n_\A = n_{\bar\A}$ & $n_\rho = n_{\A\rho} = n_{\A^2\rho}$ & $n_{\rm P}$
\\\hline\hline
\rowcolor{Blue}
2 & 1 & 0 & 2 & 3 & 15
\\
\rowcolor{Blue}
4 & 1 & 1 & 1 & 3 & 15
\\
4 & 2 & 0 & 4 & 6 & 30
\\
\rowcolor{Blue}
6 & 1 & 2 & 0 & 3 & 15
\\
6 & 2 & 1 & 3 & 6 & 30
\\
6 & 3 & 0 & 6 & 9 & 45
\\\hline
\end{tabular}
\caption{Possible numbers of point-like operators that satisfy the torus one-point modular invariance \eqref{RhoT1Pt} and \eqref{AlphaT1Pt}. Here $n_{\rm V}$ denotes the total number of vacua (local operators), comprised of $n_{\bf r}$ copies of representation $\bf r$, where ${\bf r} = +, \, -, \, {\bf 2}$; $n_\cL$ denotes the number of defect operators in {\it each} $\cL$, for $\cL = \A, \, \bar\A, \, \rho, \, \A\rho, \, \A^2\rho$; $n_{\rm P}$ denotes the total number of point-like (local and defect) operators.  Only the highlighted cases with $n_{\bf 1} = 1 , \, n_\rho = 3, \, n_{\rm P} = 15$ are considered in this note.
}
\label{Tab:Possibilities}
\end{table}

Consider the modular invariance of the torus partition function with the invertible TDL $\A$ wrapped around a one-cycle
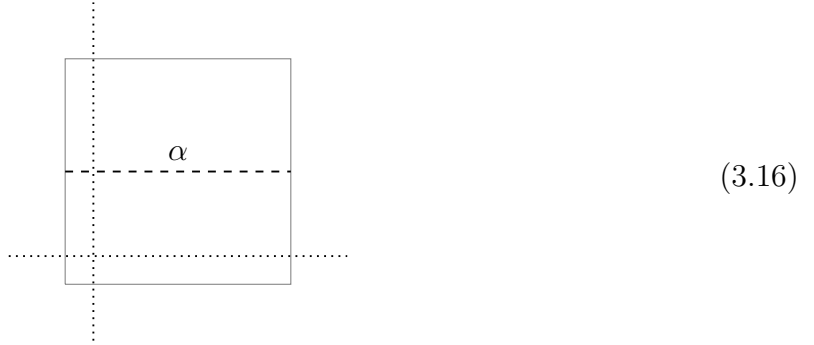
\ie
\label{AlphaT1Pt}
\begin{gathered}
\begin{tikzpicture}[scale = 1.5]
\draw [bg] (0,0) -- (2,0) -- (2,2) -- (0,2) -- (0,0);
\draw [line,dashed] (0,1) -- (1,1) node[above] {$\A$} -- (2,1);
\draw [line,dotted] (-.5,.25)--(2.5,.25);
\draw [line,dotted] (.25,-.5)--(.25,2.5);
\end{tikzpicture}
\end{gathered}
\fe
Modular invariance requires
\ie
{\rm Tr}_{\cal H} \, \wh\A = {\rm Tr}_{{\cal H}_\A} 1 \in \bZ_{\ge0} \, .
\fe
Hence the $\A$ TDL hosts
\ie
n_\A = 2 n_{\bf 1}
- n_{\bf 2}
\fe
defect operators.  The total number of point-like operators is
\ie
n_{\rm P} \equiv g + 2 n_\A + 3 n_\rho
= (2 n_{\bf 1} + 2 n_{\bf 2}) + 2 (2 n_{\bf 1} - n_{\bf 2}) + 9 n_{\bf 1}
= 15 n_{\bf 1} \, .
\fe

The first few possibilities are listed in Table~\ref{Tab:Possibilities} in the order of increasing $n_{\rm V}$.  Whenever $n_{\bf 2} = 0$, the $\bZ_3$ symmetry is not faithfully realized on the vacua.  In the following, we consider the three minimal cases totaling $n_{\rm P} = 15$ point-like operators, highlighted in Table~\ref{Tab:Possibilities}; each case has $n_{\bf 1} = 1$ and $n_\rho = 3$.  Eventually we will succeed in constructing a TFT realizing $n_{\rm V} = 6$, but along the way we also derive various constraints on $n_{\rm V} = 2, \, 4$.

\section{Transparency and $\bZ_3$ symmetry}
\label{Sec:Z3}

This note works in a gauge of the $\cH_3$ fusion category that manifests its ``transparent'' property \cite{Huang:2020lox}---the associator involving any invertible topological defect line (TDL) is the identity morphism.  In terms of the $F$-symbols, it means that every $F$-symbol with an external invertible TDL takes value one.  Hence invertible TDLs can be attached or detached ``freely'', changing the isomorphism classes of other involved TDLs but without generating extra $F$-symbols.  Several diagrammatic identities are illustrated below:
\ie
\label{DiagrammaticIdentities}
&
(a) \hspace{.25in}
\begin{gathered}
\begin{tikzpicture}[scale = 1]
\draw (0,0.8) node[left=-2pt] {$\rho_i$};
\draw (0,0) node[left=-2pt] {$\rho_{i+1}$};
\draw (0,-0.8) node[left=-2pt] {$\rho_i$};
\draw [line] (0,-1) -- (0,1);
\draw [line,dashed,-<-=.5] (0,0) ++(-90:.5) arc (-90:90:.5);
\end{tikzpicture}
\end{gathered} 
~~=~~
\begin{gathered}
\begin{tikzpicture}[scale = 1]
\draw (0,0) node[left=-2pt] {$\rho_i$};
\draw [line] (0,-1) -- (0,1);
\end{tikzpicture}
\end{gathered} 
\hspace{.5in}
(b) \hspace{.25in}
\begin{gathered}
\begin{tikzpicture}[scale = 1]
\draw (0,0.8) node[left=-2pt] {$\rho_i$};
\draw (0,0) node[left=-2pt] {$\rho_{i+1}$};
\draw (0,-0.8) node[left=-2pt] {$\rho_i$};
\draw [line] (0,-1) -- (0,1);
\draw [line,dashed,->-=.7] (0,1) ++(-90:.5) arc (-90:0:.5);
\draw [line,dashed,-<-=.7] (0,-1) ++(90:.5) arc (90:0:.5);
\end{tikzpicture}
\end{gathered} 
~~=~~
\begin{gathered}
\begin{tikzpicture}[scale = 1]
\draw (0,0) node[left=-2pt] {$\rho_i$};
\draw [line] (0,-1) -- (0,1);
\draw [line,dashed,->-=.5] (.5,-1) -- (.5,1);
\end{tikzpicture}
\end{gathered} 
\hspace{.5in}
\\
& (c) \hspace{.25in}
\begin{gathered}
\begin{tikzpicture}[scale = 1]
\node at (0,0.8)[left=-2pt] {$\rho_i$};
\node at (0,0)[left=-2pt] {$\rho_{i+1}$};
\node at (0,-0.8)[left=-2pt] {$\rho_{i+2}$};
\draw [line] (0,-1) -- (0,1);
\draw [line,dashed,->-=.5] (0,-.5) -- (.5,0);
\draw [line,dashed,->-=.5] (0,.5) -- (.5,0);
\draw [line,dashed,->-=.7] (1,0) -- (.5,0);
\end{tikzpicture}
\end{gathered}
~~=~~
\begin{gathered}
\begin{tikzpicture}[scale = 1]
\node at (0,0.8)[left=-2pt] {$\rho_i$};
\node at (0,-0.8)[left=-2pt] {$\rho_{i+2}$};
\draw [line] (0,-1) -- (0,1);
\draw [line,dashed,->-=.6] (1,0) -- (0,0);
\end{tikzpicture}
\end{gathered}
\hspace{.5in}
(d) \hspace{.25in}
\begin{gathered}
\begin{tikzpicture}[scale = 1]
\node at (-0.8,0.2) {$\rho_i$};
\draw [line] (0,0) -- (-1,0);
\draw [line] (0,0) -- (.25,-.435) node[left=-2pt] {$\rho_k$} -- (.5,-.87) node[above right=-3pt] {$\rho_{k+1}$};
\draw [line] (0,0) -- (.25,.435) node[left=-2pt] {$\rho_j$} -- (.5,.87) node[below right=-3pt] {$\rho_{j+1}$};
\draw [line,dashed,->-=.5] (.33,-.58) -- (.33,.58);
\end{tikzpicture}
\end{gathered}
~~=~~
\begin{gathered}
\begin{tikzpicture}[scale = 1]
\node at (-0.8,0.2) {$\rho_i$};
\draw [line] (0,0) -- (-1,0);
\draw [line] (0,0) -- (.5,-.87) node[above right=-3pt] {$\rho_{k+1}$};
\draw [line] (0,0) -- (.5,.87) node[below right=-3pt] {$\rho_{j+1}$};
\end{tikzpicture}
\end{gathered}
\\
&
(e) \hspace{.25in}
\begin{gathered}
\begin{tikzpicture}[scale = 1]
\node at (0,0.8)[left=-2pt]  {$\rho_i$};
\node at (0,-0.8)[left=-2pt]  {$\rho_{i+1}$};
\draw [line] (0,-1) -- (0,1);
\draw [line,dashed,->-=1] (1,0) -- (-1,0);
\end{tikzpicture}
\end{gathered}
~~=~~
\begin{gathered}
\begin{tikzpicture}[scale = 1]
\node at (0,0.8)[left=-2pt]  {$\rho_i$};
\node at (0,0)[left=-2pt]  {$\rho_{i-1}$};
\node at (0,-0.8)[left=-2pt]  {$\rho_{i+1}$};
\draw [line] (0,-1) -- (0,1);
\draw [line,dashed,->-=.6] (1,.4) -- (0,.4);
\draw [line,dashed,-<-=.5] (-1,-.4) -- (0,-.4);
\end{tikzpicture}
\end{gathered}
~~=~~
\begin{gathered}
\begin{tikzpicture}[scale = 1]
\node at (0,0.8)[left=-2pt] {$\rho_i$};
\node at (0,0)[left=-2pt] {$\rho_{i-1}$};
\node at (0,-0.8)[left=-2pt] {$\rho_{i+1}$};
\draw [line] (0,-1) -- (0,1);
\draw [line,dashed,->-=.6] (1,-.4) -- (0,-.4);
\draw [line,dashed,-<-=.5] (-1,.4) -- (0,.4);
\end{tikzpicture}
\end{gathered}
\\
&
(f) \hspace{.25in}
\begin{gathered}
\begin{tikzpicture}[scale = 1]
\node at (0,0.8)[left=-2pt] {$\rho_i$};
\node at (0,-0.8)[left=-2pt] {$\rho_i$};
\draw [line] (0,-1) -- (0,1);
\draw [line,dashed,-<-=.5] (1,0) -- (0,0);
\draw [line,dashed,-<-=.5] (-1,0) -- (0,0);
\end{tikzpicture}
\end{gathered}
~~=~~
\begin{gathered}
\begin{tikzpicture}[scale = 1]
\node at (0,0.8)[left=-2pt] {$\rho_i$};
\node at (0,0)[left=-2pt] {$\rho_{i+1}$};
\node at (0,-0.8)[left=-2pt] {$\rho_i$};
\draw [line] (0,-1) -- (0,1);
\draw [line,dashed,-<-=.5] (1,.4) -- (0,.4);
\draw [line,dashed,-<-=.5] (-1,-.4) -- (0,-.4);
\end{tikzpicture}
\end{gathered}
~~=~~
\begin{gathered}
\begin{tikzpicture}[scale = 1]
\node at (0,0.8)[left=-2pt] {$\rho_i$};
\node at (0,0)[left=-2pt] {$\rho_{i-1}$};
\node at (0,-0.8)[left=-2pt] {$\rho_i$};
\draw [line] (0,-1) -- (0,1);
\draw [line,dashed,-<-=.5] (1,-.4) -- (0,-.4);
\draw [line,dashed,-<-=.5] (-1,.4) -- (0,.4);
\end{tikzpicture}
\end{gathered}
\fe
Importantly, the four-way junctions in $(e)$ and $(f)$ are unambiguously defined.

In \cite{Huang:2020lox}, transparency and the $\bZ_3$ symmetry were exploited to reduce the pentagon identity so that the $F$-symbols could be efficiently solved.  Below, in attempting to construct a topological field theory, the utilization of the $\bZ_3$ symmetry is also essential in reducing the amount of independent data.

\subsection{$\bZ_3$ relations for lassos and dumbbells}

Let $\cO_q$ be a local operator with $\bZ_3$-charge $q \in \{0, \, \pm 1\}$, and consider the lasso
\ie
\bell{0}{$\rho_j$}{$\rho_i$}{$\cO_q$} \, .
\fe
The $\bZ_3$ symmetry relates lassos with different triples $(q, i, j)$ as follows: replace $\cO_q$ using the equalities
\ie
\cO_q
~=~
\omega^q \
\begin{gathered}
\begin{tikzpicture}[scale = .5]
\draw [line] (0,0) \dt{below}{0}{$\cO_q$};
\draw [line,dashed,->-=.25] (0,0) circle (1.5);
\end{tikzpicture}
\end{gathered}
~=~
\omega^{-q} \
\begin{gathered}
\begin{tikzpicture}[scale = .5]
\draw [line] (0,0) \dt{below}{0}{$\cO_q$};
\draw [line,dashed,-<-=.25] (0,0) circle (1.5);
\end{tikzpicture}
\end{gathered}
\fe
and fuse the $\bZ_3$ symmetry line with $\rho_i$ (apply \eqref{DiagrammaticIdentities}$(b)$ and then $(d)$) to obtain the relations
\ie
\label{lassoXi}
& \bell{0}{$\rho_j$}{$\rho_i$}{$\cO_q$} = \omega^q \, \bell{0}{$\rho_j$}{$\rho_{i-1}$}{$\cO_q$} = \omega^{-q} \,  \bell{0}{$\rho_j$}{$\rho_{i+1}$}{$\cO_q$} \, .
\fe

Next consider the dumbbell
\ie
\dum{.5}{$\rho_i$}{$\rho_j$}{$\rho_k$}{}{} \, ,
\fe
where each empty dot denotes an arbitrary local operator insertion.  The $\bZ_3$ action on the dumbbell (circling it with a clockwise $\bZ_3$ loop) gives (see \eqref{DiagrammaticIdentities}$(e)$ for the meaning of the four-way junction)
\ie
\label{dumbXi}
\begin{gathered}
\begin{tikzpicture}[scale = .5]
\dumpre{.5}{$\rho_i$}{$\rho_j$}{$\rho_k$}{}{}
\draw [line,dashed,-<-=.25] (0,0) ellipse (5.5 and 3);
\end{tikzpicture}
\end{gathered}
\quad &= \quad
\begin{gathered}
\begin{tikzpicture}[scale = .5]
\dumpre{.5}{$\rho_i$}{$\rho_j$}{$\rho_k$}{}{}
\draw [line,dashed,-<-=1] (3,0) circle (2.5);
\draw [line,dashed,-<-=.5] (-3,0) circle (2.5);
\end{tikzpicture}
\end{gathered}
\\
\quad &= \quad
\dum{.5}{$\rho_{i-1}$}{$\rho_{j+1}$}{$\rho_{k-1}$}{}{}
\fe
Combining \eqref{lassoXi} and \eqref{dumbXi}, we obtain an identity that leaves the side loops intact and only changes the handle
\ie
\label{dumbXi2}
\begin{gathered}
\begin{tikzpicture}[scale = .5]
\dumpre{.5}{$\rho_i$}{$\rho_j$}{$\rho_k$}{$\cO_{q_1}$}{$\cO_{q_2}$}
\draw [line,dashed,-<-=.25] (0,0) ellipse (5.5 and 3);
\end{tikzpicture}
\end{gathered}
\quad = \quad
\omega^{-q_1-q_2} \dum{.5}{$\rho_i$}{$\rho_{j+1}$}{$\rho_k$}{$\cO_{q_1}$}{$\cO_{q_2}$} \, ,
\fe
which will prove useful in Section~\ref{Sec:RhoAction}.  A mnemonic is that the $\bZ_3$ symmetry line measures the {\it opposite} $\bZ_3$-charge of the local operators $\cO_{q_1}$ and $\cO_{q_2}$ placed inside a dumbbell, because the $\bZ_3$ symmetry line changes orientation when it crosses a $\rho_i$ TDL, as illustrated in \eqref{DiagrammaticIdentities}$(f)$.

\subsection{$\bZ_3$ action on defect operators}
\label{Sec:Z3Defect}

Recall that each $\rho_i$ TDL hosts three independent defect operators.  We work in an orthonormal basis and denote them by
\ie
o_{iA} \, , \quad i = 0, 1, 2 \, , \quad A = 1, 2, 3 \, ,
\qquad \text{with} \quad
\la o_{iA} o_{jB} \ra = \D_{ij} \D_{AB} \, .
\fe
Note that there is still an ${\rm O}(3)^3$ basis freedom.  The $\bZ_3$ action on a defect operator $o_i \in {\cal H}_{\rho_i}$ is defined by the lasso (see \eqref{DiagrammaticIdentities}$(e)$ for the meaning of the four-way junction)
\ie
\label{Z3Lasso}
\begin{gathered}
\begin{tikzpicture}[scale = 1]
\draw [line] (0,-.5) node[below=-2pt] {$\rho_{i+1}$} -- (0,.5) \dt{above}{-2}{$o_{iA}$};
\draw [line,dashed,-<-=.5] (0,.6) circle (.5);
\end{tikzpicture}
\end{gathered} 
\quad = \quad
\begin{gathered} 
\begin{tikzpicture}[scale=.5]
\draw [bg] (0,0) ++ (180:2 and 1) arc (180:360:2 and 1);
\draw [bg,dashed] (0,0) ++ (0:2 and 1) arc (0:180:2 and 1);
\draw [bg] (0,4) ellipse (2 and 1);
\draw [bg] (-2,0) -- (-2,4);
\draw [bg] (2,0) -- (2,4);
\draw [line] (0,-1) node[below] {$|o_{iA}\rangle$} -- node[right=-2pt] {$\rho_i$} (0,1) -- node[right=-2pt] {$\rho_{i+1}$} (0,3);
\draw [line,dashed,-<-=.3] (0,2) ++ (180:2 and 1) arc (180:360:2 and 1);
\draw [line,dashed,opacity=.5] (0,2) ++ (0:2 and 1) arc (0:180:2 and 1);
\end{tikzpicture}
\end{gathered} 
\quad = \quad 
\begin{gathered}
\begin{tikzpicture}[scale = 1.5]
\draw [bg] (0,0) -- (2,0) -- (2,2) -- (0,2) -- (0,0);
\draw [line] (1,0) node[below] {$|o_{iA}\rangle$} -- (1,.5) node[right=-2pt] {$\rho_i$} -- (1,1.5) node[right=-2pt] {$\rho_{i+1}$} -- (1,2);
\draw [line,dashed,->-=.5] (1,.9) -- (2,1);
\draw [line,dashed,->-=.5] (0,1) -- (1,1.1);
\end{tikzpicture}
\end{gathered}
\, ,
\fe
where in the last diagram, the left and right edges of the square are identified to represent a cylinder.  Performing the $\bZ_3$ action three times on ${\cal H}_{\rho_i}$ becomes a trivial action, as illustrated by the sequence of $F$-moves
\ie
\begin{gathered}
\begin{tikzpicture}[scale = 1.5]
\draw [bg] (0,0) -- (2,0) -- (2,2) -- (0,2) -- (0,0);
\draw [line] (1,0) -- (1,2);
\draw [line,dashed,->-=.5] (1,.4) -- (2,.5);
\draw [line,dashed,->-=.5] (1,.9) -- (2,1);
\draw [line,dashed,->-=.5] (1,1.4) -- (2,1.5);
\draw [line,dashed,->-=.5] (0,.5) -- (1,.6);
\draw [line,dashed,->-=.5] (0,1) -- (1,1.1);
\draw [line,dashed,->-=.5] (0,1.5) -- (1,1.6);
\end{tikzpicture}
\end{gathered}
\quad = \quad
\begin{gathered}
\begin{tikzpicture}[scale = 1.5]
\draw [bg] (0,0) -- (2,0) -- (2,2) -- (0,2) -- (0,0);
\draw [line] (1,0) -- (1,2);
\draw [line,dashed,->-=.5] (1,.9) -- (1.5,.9);
\draw [line,dashed,->-=.3,-<-=.8] (1,.9) ++(-90:.5) arc (-90:90:.5);
\draw [line,dashed,->-=.5] (.5,1.1) -- (1,1.1);
\draw [line,dashed,-<-=.3,->-=.8] (1,1.1) ++(270:.5) arc (270:90:.5);
\end{tikzpicture}
\end{gathered}
\quad = \quad
\begin{gathered}
\begin{tikzpicture}[scale = 1.5]
\draw [bg] (0,0) -- (2,0) -- (2,2) -- (0,2) -- (0,0);
\draw [line] (1,0) -- (1,2);
\end{tikzpicture}
\end{gathered} \, .
\fe
We make use of the ${\rm O}(3)^2 \subset {\rm O}(3)^3$ basis freedom such that the lasso \eqref{Z3Lasso} representing the $\bZ_3$ action takes
\ie
\bZ_3: \quad 
o_{1A} \to o_{2A} \to o_{3A} \to o_{1A} \, .
\fe

The $\bZ_3$ action also gives rise to relations among the dynamical data.  For instance, consider the $\bZ_3$ action on the operator product of $o_{iA}$ and $o_{iB}$
\ie
&
\begin{gathered}
\begin{tikzpicture}[scale = 1]
\draw [line,dashed,-<-=.5] (0,0) circle (1);
\draw [line] (0,-.5) \dt{below}{-2}{$o_{iB}$} -- (0,.5) \dt{above}{-2}{$o_{iA}$};
\end{tikzpicture}
\end{gathered}
\quad = \quad
\begin{gathered}
\begin{tikzpicture}[scale = 1]
\draw [line] (0,-.5) \dt{below}{-2}{$o_{iA}$} -- (0,.5) \dt{above}{-2}{$o_{iB}$};
\draw [line,dashed,-<-=.5] (0,.6) circle (.5);
\draw [line,dashed,-<-=.5] (0,-.6) circle (.5);
\end{tikzpicture}
\end{gathered} \, .
\fe
If the vacuum expectation value is taken, possibly in the presence of other local operators, the $\bZ_3$ symmetry line can be deformed to shrink in some other patch while picking up the $\bZ_3$-charges of other local operators.  This process gives rise to identities among correlators.  Similarly, the $\bZ_3$ action
\ie
\label{Z3Action3Pt}
\begin{gathered}
\begin{tikzpicture}[scale = 1]
\draw [line,dashed,-<-=.5] (0,0) circle (2);
\draw [line] (0,0) -- (-1,0) \dt{left}{-4}{$o_{iA}$};
\draw [line] (0,0) -- (.5,-.87) \dt{below right}{-5}{$o_{jB}$};
\draw [line] (0,0) -- (.5,.87) \dt{above right}{-5}{$o_{kC}$};
\end{tikzpicture}
\end{gathered}
\quad = \quad
\begin{gathered}
\begin{tikzpicture}[scale = 2]
\draw [line] (0,0) -- (-1,0) \dt{left}{-4}{$o_{iA}$};
\draw [line,dashed,-<-=.5] (-1,0) circle (.4);
\draw [line] (0,0) -- (.5,-.87) \dt{below right}{-5}{$o_{jB}$};
\draw [line,dashed,-<-=.5] (.5,-.87) circle (.4);
\draw [line] (0,0) -- (.5,.87) \dt{above right}{-5}{$o_{kC}$};
\draw [line,dashed,-<-=.5] (.5,.87) circle (.4);
\draw [line,dashed,->-=.5] (-.5,0) -- (.17,.29);
\draw [line,dashed,->-=.5] (.25,.43) -- (.17,-.29);
\draw [line,dashed,->-=.5] (.25,-.43) -- (-.33,0);
\end{tikzpicture}
\end{gathered}
\quad = \quad
\begin{gathered}
\begin{tikzpicture}[scale = 1]
\draw [line] (0,0) -- (-1,0) \dt{left}{-4}{$o_{iA}$};
\draw [line,dashed,-<-=.5] (-1,0) circle (.8);
\draw [line] (0,0) -- (.5,-.87) \dt{below right}{-5}{$o_{jB}$};
\draw [line,dashed,-<-=.5] (.5,-.87) circle (.8);
\draw [line] (0,0) -- (.5,.87) \dt{above right}{-5}{$o_{kC}$};
\draw [line,dashed,-<-=.5] (.5,.87) circle (.8);
\end{tikzpicture}
\end{gathered}
\fe
implies identities among different three-point coefficients, when the sphere vacuum expectation value is taken.

We can nucleate $\bZ_3$ loops inside or outside a lasso to change the species of the $\rho_i$ TDLs, resulting in the relations
\ie
\label{LassoRelations}
& \bell{1}{$o_{jB}$}{$\rho_i$}{$\cO_q$}
~=~
\bell{1}{$o_{j+1,B}$}{$\rho_{i-1}$}{$\cO_q$}
~=~
\omega^q \, \bell{1}{$o_{jB}$}{$\rho_{i-1}$}{$\cO_q$}
\, ,
\\
&
\lassoX{1}{$o_{jB}$}{$\rho_k$}{$\rho_\ell$}{$o_{iA}$} 
~=~
\lassoX{1}{$o_{j+1,B}$}{$\rho_{k-1}$}{$\rho_{\ell-1}$}{$o_{iA}$}
~=~
\lassoX{1}{$o_{jB}$}{$\rho_{k-1}$}{$\rho_{\ell-1}$}{$o_{i+1,A}$} \, .
\fe

\section{Bootstrap constraints}
\label{Sec:Consistency}

Given the spectral constraints derived in Section~\ref{Sec:Modular}, our goal now is to solve for a minimal defect topological field theory (TFT) with a total of $n_{\rm P} = 15$ point-like operators, and the number of vacua (local operators) can be $n_{\rm V} = 2, \, 4, \, 6$.  For each case, there is one nontrivial $\bZ_3$-neutral local operator $v$ and three defect operators $o_{iA}$ on each $\rho_i$ line.  The remaining four point-like operators can be two pairs of $\bZ_3$-charged local operators $u_a, \, \bar u_a$, two pairs of $\bZ_3$ defect operators $w_a \in {\cal H}_\A, \, \bar w_a \in {\cal H}_{\A^2}$, or a pair of each.

In this section, we delineate constraints of crossing symmetry and modular invariance that were formulated in generality in Section~\ref{Sec:Crossing}.  For simplicity, we ignore the constraints involving $\bZ_3$ defect operators $w_a, \, \bar w_a$, and only consider the part of crossing symmetry that is equivalent to the associativity involving at least one local operator.  More general crossing symmetry is deferred to Appendix~\ref{App:Defect}.  

We reserve the $i = 0, 1, 2$ index for the species of the $\rho_i$ line, the $A = 1, 2, 3$ index for the species of the defect operators of each $\rho_i$ line, and the $a = 1, \dotsc, n_{\bf 2}$ index for the species of $\bZ_3$-charged local operators.  Note that the $\bZ_3$-charged operators $u_a, \, \bar u_a$ have a ${\rm U}(n_{\bf 2})$ basis freedom.

\subsection{Local operator algebra and associativity}
\label{Sec:LocalAssociativity}

The most general local operator algebra consistent with the $\bZ_3$ symmetry is
\ie
& v \times v = 1 + \B v \, , \quad v \times u_a = \sum_b \xi_{ab} u_b \, ,
\\
& u_a \times \bar u_b = \D_{ab} + \xi_{ab} v \, , \quad u_a \times u_b = \sum_c \sigma_{abc} \bar u_c \, . 
\fe
The following are the constraints from associativity.
\begin{itemize}

\item
$\underline{u_a u_b u_c}$
\ie
& \sigma_{abc} = \sigma_{bca} \, , \quad \sum_d \sigma_{abd} \, \xi_{ed} = \sum_d \sigma_{bcd} \, \xi_{ad} \, , 
\\
& \sum_e \sigma_{abe} \sigma_{cde} = \sum_e \sigma_{ade} \sigma_{bce} = \sum_e \sigma_{ace} \sigma_{bde} \, .
\fe
Hence $\sigma_{abc}$ is totally symmetric.

\item
$\underline{u_a \bar u_b v}$
\ie
\xi_{ab} = {\bar\xi}_{ba} \, , \quad \D_{ab} + \B \xi_{ab} = \sum_c \xi_{ac} \, {\bar\xi}_{bc} = \sum_c {\bar\xi}_{bc} \, \xi_{ac} \, ,
\fe
The first condition says that $\xi_{ab}$ is Hermitian, which allows us to use the ${\rm U}(n_{\bf 2})$ basis freedom to diagonalize $\xi_{ab}$.  Then the second condition, which also encompasses the associativity of $\underline{u_a v v}$, is solved by
\ie
\label{tau}
\xi_{ab} = \xi_a \D_{ab} \, , \quad \xi_a = {\B \pm \sqrt{\B^2 + 4} \over 2} \, .
\fe

\item
$\underline{u_a u_b v}$
\ie
\label{uuv}
\sum_c \sigma_{abc} {\bar\xi}^{cd} = \sum_c \xi_{bc} \sigma_{acd} = \sum_c \xi_{ac} \sigma_{bcd} \, .
\fe

\item
$\underline{u_a u_b \bar u_c}$
\ie
\label{uuubar}
\sum_d \sigma_{abd} \bar\sigma_{dce} = \D_{bc} \D_{ae} + \xi_{bc} \xi_{ae} \, .
\fe
In the special case of $a = e$ and $b = c$,
\ie
\label{uuubar2}
\sum_d \sigma_{abd} \bar\sigma_{dba} = 1 + \xi_a \xi_b \, .
\fe

\end{itemize}

\subsection{Mixed local and $\rho$ defect operators}
\label{Sec:DefectAssociativity}

The most general operator algebra involving mixed local and $\rho$ defect operators is
\ie
\label{DefectOP}
& \defect{1}{0}{$\rho_i$}{$o_{iA}$}{$o_{iB}$} = \D_{AB} + \kappa^i_{AB} v + \sum_a \left( \bar\lambda^i_{AB;a} u_a + \lambda^i_{AB;a} \bar u_a \right) \, ,
\\
& \defect{0}{0}{$\rho_i$}{}{$o_{iA}$} v = \sum_B \kappa^i_{AB} \defect{0}{0}{$\rho_i$}{}{$o_{iB}$} \, ,
\quad
\defect{0}{0}{$\rho_i$}{}{$o_{iA}$} u_a = \sum_B \lambda^i_{AB;a} \defect{0}{0}{$\rho_i$}{}{$o_{iB}$} \, ,
\fe
where $\kappa^i_{AB}$ and $\lambda^i_{AB;a}$ are both symmetric in $A, B$, and the $\bZ_3$ action \eqref{Z3Action3Pt} implies that
\ie
\label{KappaLambda}
\kappa^{i+1}_{AB} = \kappa^i_{AB} \, , 
\quad
\lambda^{i+1}_{AB;a} = \omega^{-1} \lambda^i_{AB;a} \, . 
\fe
The following are the constraints from associativity.
\begin{itemize}

\item 
$\underline{o_{iA} o_{iB} v}$
\ie
& \defect{1}{0}{$\rho_i$}{$o_{iA}$}{$o_{iB}$} v 
\\
&= \D_{AB} v + \kappa^i_{AB} (1 + \B v) + \sum_{a,b} \left( \bar\lambda^i_{AB;b} \xi_{ba} u_a + \lambda^i_{AB;b} {\bar\xi}_{ba} \bar u_a \right)
\\
&= \kappa^i_{AB} + \left( \sum_{C} \kappa^i_{AC} \kappa^i_{BC} \right) v
+ \sum_{C} \kappa^i_{A{C}} \sum_a ( \bar\lambda^i_{B{C};a} u_a + \lambda^i_{B{C};a} \bar u_a) \, .
\fe
Hence,
\ie
\label{oovv}
& \sum_{C} \kappa^i_{AC} \kappa^i_{BC} = \D_{AB} + \B \kappa^i_{AB} \, ,
\fe
which also encompasses the associativity of $\underline{o_{iA} v v}$,
and
\ie
\label{oovu}
& \sum_{C} \kappa^i_{AC} \lambda^i_{BC;a} = \sum_b \lambda^i_{AB;b} {\bar\xi}_{ba} \, .
\fe

\item 
$\underline{o_{iA} o_{iB} u_a}$
\ie
& \defect{1}{0}{$\rho_i$}{$o_{iA}$}{$o_{iB}$} u_a
\\
&= \D_{AB} u_a + \sum_b \kappa^i_{AB} \xi_{ab} u_b + \sum_{b,c} \bar\lambda^i_{AB;b} \sigma_{abc} \bar u_c + \lambda^i_{AB;a} + \sum_b \lambda^i_{AB;b} \, \xi_{ab} v
\\
&= \lambda^i_{AB;a} + \left( \sum_{C} \lambda^i_{AC;a} \kappa^i_{BC} \right) v
+ \sum_{C} \lambda^i_{A{C};a} \sum_b ( \bar\lambda^i_{B{C};b} u_b + \lambda^i_{B{C};b} \bar u_b) \, .
\fe
Hence,
\ie
\label{oouu}
& \sum_C \lambda^i_{AC;a} \bar\lambda^i_{BC;b} = \D_{AB} \D_{ab} + \kappa^i_{AB} \xi_{ab} \, ,
\quad
\sum_C \lambda^i_{AC;a} \lambda^i_{BC;b} = \sum_c \bar\lambda^i_{AB;c} \sigma_{abc} \, .
\fe

\end{itemize}

\subsection{$\rho$ action on local operators}
\label{Sec:RhoAction}

Let us study the analog of charge conservation \eqref{ChargeConservation} for the non-invertible TDLs $\rho_i$.  We will constrain the $\rho_i$ action on local operators,
\ie
& \x{.5}{$\rho_i$}{1} = \zeta \, , \quad \x{.5}{$\rho_i$}{$v$} = - \zeta^{-1} v \, ,
\quad
\x{.5}{$\rho_i$}{$u_a$} = \omega^i \sum_b R_{ab} \bar u_b \, ,
\fe
and the lassos on local operators,
\ie
\label{xiv}
\varepsilon^i_A \equiv \bell{1}{$o_{iA}$}{$\rho_i$}{$v$} \, , 
\quad
\C^i_{aA} \equiv \bell{1}{$o_{iA}$}{$\rho_i$}{$u_a$} \, .
\fe
The $\bZ_3$ action relations \eqref{LassoRelations} imply that
\ie
\label{EpsilonGammaZ3}
\varepsilon^{i+1}_A = 
\omega^{-i} 
\varepsilon^i_A \, , 
\quad
\C^{i+1}_{aA} = 
\omega^{-i} 
\C^i_{aA} \, .
\fe
Note that in writing \eqref{tau}, we already used the ${\rm U}(n_{\bf 2})$ freedom to diagonalize the operator product $u_a \, \bar u_b$, so we can no longer use it to simplify $R_{ab}$.  In the following, we make frequent use of the explicit values of the $F$-symbols
\ie
(F^{\rho_i,\rho_i,\rho_i}_{\rho_i})_{\cI, \cI} = \zeta^{-1} \, , \quad (F^{\rho_i,\rho_i,\rho_i}_{\rho_i})_{\cI, \rho_j} = \zeta^{-1} \, , \quad \zeta \equiv {3 + \sqrt{13} \over 2}
\fe
from \eqref{FSymbolsAlt}.

First, let us revisit the requirement that $u_a$ transforms as a representation of the fusion ring.\footnote{The representation given in \eqref{Rep2} was specialized to a particular basis for $u_a$.  Here we prioritize the use of the ${\rm U}(n_{\bf 2})$ basis freedom to diagonalize $\xi_{ab}$ in \eqref{tau}, so the requirement that $u_a$ transforms as a representation needs to be rewritten in a basis-independent fashion.
}
The consideration of
\ie
\begin{gathered}
	\begin{tikzpicture}[scale = .75]
		\draw (2,0) node[right=-2pt] {$\rho$};
		\draw (1,0) node[right=-2pt] {$\rho$};
		\draw [line] (0,0) \dt{below}{0}{$u_a$} circle (1) circle (2);
	\end{tikzpicture}
\end{gathered}
\quad = \quad
u_a + \sum_i\x{.5}{$\rho_i$}{$u_a$}
\fe
leads to a constraint
\ie
\label{XiBarXi}
\sum_c R_{ac} \overline{R}_{cb} = \D_{ab} + \sum_i \omega^i R_{ab} = \D_{ab} \, ,
\fe
where the left side comes from shrinking the inner and outer $\rho$ loops consecutively, and the right side from fusing them before shrinking.\footnote{The fusion of the two $\rho$ TDLs can be understood as an $F$-move followed by the shrinking of the $\rho$ loop.
}

Now, following the $\downarrow$ direction in \eqref{ChargeConservation}, we circle $\rho_i$ on the operator product of local operators, and apply the $F$-move to obtain
\ie
\begin{gathered}
	\begin{tikzpicture}[scale = .5]
		\node at (4.5,0) {$\rho_i$};
		\draw [line] (-2,0) \dt{below}{0}{$\cO_{q_1}$};
		\draw [line] (2,0) \dt{below}{0}{$\cO_{q_2}$};
		\draw [line, dashed] (0,-2.5) -- (0,0) node[right] {$\cI$} -- (0,2.5);
	    \draw [line] (0,0) ellipse (4cm and 2.5cm);
	\end{tikzpicture}
\end{gathered}
=
\sum_{j'} \dum{.5}{$\rho_i$}{$\rho_{j'}$}{$\rho_i$}{$\cO_{q_1}$}{$\cO_{q_2}$} \, .
\fe
Using the $\bZ_3$ action \eqref{dumbXi2}, we can simplify the sum of dumbbells to 
\ie
3 \left. \dum{.5}{$\rho_i$}{$\rho_j$}{$\rho_i$}{$\cO_{q_1}$}{$\cO_{q_2}$} \right|_\text{$\bZ_3$-charge $- (q_1+q_2)$} \, ,
\fe
where $j$ is arbitrary.  We might as well set $j = i$.  In the following, we equate the above to the $\rightarrow\,\downarrow$ direction of \eqref{ChargeConservation} where the local operator product is taken first.

\begin{itemize}

\item
$\underline{v \times v = 1 + \B v}$
\ie
\label{xiv2}
\zeta^{-3} \, (1+\B v) + 3 \, \zeta^{-1}
\left. \dum{.5}{$\rho_i$}{$\rho_i$}{$\rho_i$}{$v$}{$v$} \right|_\text{$\bZ_3$-neutral} = \zeta - \zeta^{-1} \B v \, .
\fe
Hence,
\ie
\label{DeltaKappa}
\sum_A (\varepsilon^i_A)^2 = \sqrt{13} \, , \quad \sum_{A,B} \varepsilon^i_A \kappa^i_{AB} \varepsilon^i_B = - {\sqrt{13} \over 3} \zeta^{-1} \B \, .
\fe

\item
$\underline{u_a \times \bar u_b = \D_{ab} + \xi_{ab} v}$
\ie
\label{xiuubar}
& \zeta^{-1} \, \left( \D_{ab} + \sum_{c,d} R_{ac}\overline{R}_{bd} \xi_{dc} v \right) + 3 \, \zeta^{-1} \left. \dum{.5}{$\rho_i$}{$\rho_i$}{$\rho_i$}{$u_a$}{$\bar u_b$} \right|_\text{$\bZ_3$-neutral}
\\
& \hspace{.5in}
= \zeta \D_{ab} - \zeta^{-1} \xi_{ab} v \, .
\fe
Hence,
\ie
\label{xiuubarResult}
\sum_A \C^i_{aA} \bar\C^i_{bA} = 
\zeta \D_{ab} \, , 
\fe
\ie
\label{xiuubarResult2}
\sum_{A,B} \C^i_{aA} \bar\C^i_{bB} \kappa^i_{AB} = - \frac13 
\left( \xi_{ab} + \sum_{c,d} R_{ac} \overline{R}_{bd} \xi_{dc} \right)  \, .
\fe

\item
$\underline{u_a \times u_b = \sum_c \sigma_{abc} \bar u_c}$
\ie
& \zeta^{-1} \omega^{-i} \sum_{d,e,f} R_{ad} R_{be} \bar\sigma_{def} u_f + 3 \, \zeta^{-1} \left.
\dum{.5}{$\rho_i$}{$\rho_i$}{$\rho_i$}{$u_a$}{$u_b$} \right|_\text{$\bZ_3$-charge 1}
\\
& \hspace{.5in} = \omega^{-i} \sum_{c,f} \sigma_{abc} \overline{R}_{cf} u_f \, .
\fe
Hence,
\ie
\label{xiuuResult}
\sum_{A,B} \C^i_{aA} \C^i_{bB} \bar\lambda^i_{AB;f} = \frac13 \omega^{-i} 
\left( \zeta \sum_c \sigma_{abc} \overline{R}_{cf} - \sum_{d,e} R_{ad} \rho_{be} \bar\sigma_{def} \right) \, .
\fe

\item
$\underline{u_a \times v = \sum_b \xi_{ab} u_b}$
\ie
\label{xiuv}
& - \zeta^{-2} \omega^{i} \sum_{b,c} R_{ab} {\bar\xi}_{bc} \bar u_c + 3 \, \zeta^{-1} \left. \dum{.5}{$\rho_i$}{$\rho_i$}{$\rho_i$}{$u_a$}{$v$} \right|_\text{$\bZ_3$-charge $-1$}
\\
& \hspace{.5in} = \omega^{i} \sum_{b,c} \xi_{ab} R_{bc} \bar u_c \, .
\fe
Hence,
\ie
\label{xiuvResult}
\sum_A \C^i_{aA} \lambda^i_{AB;c} \varepsilon^i_B = {1\over3} \omega^{i} \sum_b \left( \zeta \xi_{ab} R_{bc} + \zeta^{-1} R_{ab} {\bar\xi}_{bc} \right) \, .
\fe

\end{itemize}

\subsection{Torus one-point modular invariance}

Consider the torus one-point modular invariance \eqref{ModularInvariance} in the special case of
\ie
\cL_2 = \cL_\cO = \cI \, , \quad \cL_3 = \ocL_4 = \cL_1 \, .
\fe

\begin{itemize}

\item
\underline{$v$ with $\bZ_3$ symmetry line}
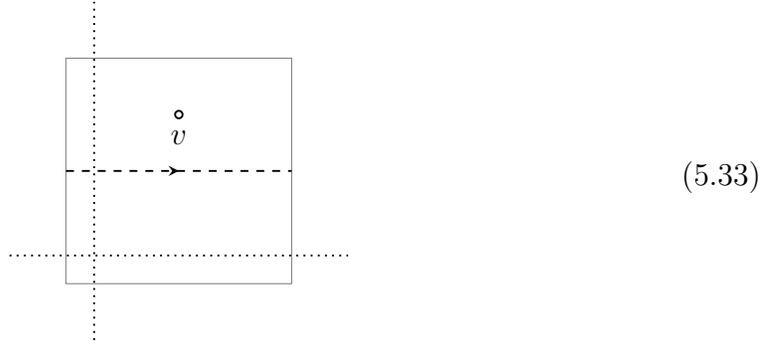
\ie
\begin{tikzpicture}[scale = 1.5]
\draw [bg] (0,0) -- (2,0) -- (2,2) -- (0,2) -- (0,0);
\draw [line,dashed,->-=.5] (0,1) -- (2,1);
\draw [line] (1,1.5) \dt{below}{0}{$v$};
\draw [line,dotted] (-.5,.25)--(2.5,.25);
\draw [line,dotted] (.25,-.5)--(.25,2.5);
\end{tikzpicture}
\fe

Let us denote the three-point coefficient of $v$ with $\bZ_3$ defect operators by
\ie
\tilde \xi_a = c(v, w_a, \bar w_a) \, .
\fe
Let us write down
\ie
\text{vertical cut} = \text{horizontal cut}
\fe
for different numbers of vacua.
\begin{enumerate}[(a)]
\item
$n_{\rm V} = 6$
\ie
\label{Z3ModularInvariance}
0 &= c(v, v, v) + \omega \sum_a c(v, u_a, \bar u_a) + \omega^2  \sum_a c(v, \bar u_a, u_a)
\\
&= c(v, v, v) - \sum_a c(v, u_a, \bar u_a)
\\
&= \B - \xi_1 - \xi_2 \, .
\fe
\item
$n_{\rm V} = 4$
\ie
\label{Z3ModularInvariance4}
\tilde\xi = \B - \xi \, .
\fe
\item
$n_{\rm V} = 2$
\ie
\label{Z3ModularInvariance2}
\tilde\xi_1 + \tilde\xi_2 = \B \, .
\fe
\end{enumerate}

\item
\underline{$v$ with $\rho$ line}
\ie
\begin{tikzpicture}[scale = 1.5]
\draw [bg] (0,0) -- (2,0) -- (2,2) -- (0,2) -- (0,0);
\draw [line] (0,1) -- (1,1) node[below] {$\rho_i$} -- (2,1);
\draw [line] (1,1.5) \dt{below}{0}{$v$};
\draw [line,dotted] (-.5,.25)--(2.5,.25);
\draw [line,dotted] (.25,-.5)--(.25,2.5);
\end{tikzpicture}
\fe
Under the vertical cut,
\ie
\sum_A c(v, o_{iA}, o_{iA}) = \sum_A \kappa^i_{AA} = {\rm tr}(\kappa^i) \, ,
\fe
and under the horizontal cut.
\ie
- \zeta^{-1} c(v, v, v) = - \zeta^{-1} \B \, ,
\fe
Hence,
\ie
\label{T1xiv}
{\rm tr}(\kappa^i)  = - \zeta^{-1} \B \, .
\fe

\item
\underline{$u_a$ with $\rho$ line}
\ie
\begin{tikzpicture}[scale = 1.5]
\draw [bg] (0,0) -- (2,0) -- (2,2) -- (0,2) -- (0,0);
\draw [line] (0,1) -- (1,1) node[below] {$\rho_i$} -- (2,1);
\draw [line] (1,1.5) \dt{below}{0}{$u_a$};
\draw [line,dotted] (-.5,.25)--(2.5,.25);
\draw [line,dotted] (.25,-.5)--(.25,2.5);
\end{tikzpicture}
\fe
Under the vertical cut
\ie
\sum_A c(u_a, o_{iA}, o_{iA}) = \sum_A \lambda^i_{AA;a} \, ,
\fe
and under the horizontal cut,
\ie
\sum_{b,c} \overline{R}_{bc} c(u_a, u_b, u_c) = \sum_{b,c} \overline{R}_{bc} \sigma_{abc} \, .
\fe
Hence,
\ie
\label{TrLambda}
\sum_A \lambda^i_{AA;a} = \sum_{b,c} \overline{R}_{bc} \sigma_{abc} \, .
\fe

\end{itemize}

\section{Topological field theory with Haagerup $\cH_3$ symmetry}
\label{Sec:TFT}

This section analyzes the bootstrap constraints delineated in the previous section.  We first narrow down the local operator algebra to a handful of possibilities, and then proceed to construct a topological field theory with six vacua realizing the Haagerup $\cH_3$ fusion category.

\subsection{Local operator algebra}
\label{Sec:LocalAlgebra}

To solve for a defect topological field theory, we begin by examining the associativity of local operators detailed in Section~\ref{Sec:LocalAssociativity}.  There we used the ${\rm U}(n_{\bf 2})$ basis freedom for $u_a$ and $\bar u_a$ to put $\xi_{ab}$ into diagonal form, and used associativity to constrain the possible eigenvalues; the result was
\ie
\label{xiSolved}
\xi_{ab} = \xi_a \D_{ab} \, , \quad \xi_a = {\B \pm \sqrt{\B^2 + 4} \over 2} \, .
\fe
In this basis, \eqref{uuv} becomes
\ie
\xi_a \, \sigma_{abc} = \xi_b \, \sigma_{abc} = \xi_c \, \sigma_{abc} \, .
\fe
Then for any pair $(a, b)$ such that $\xi_a \neq \xi_b$, it follows that $\sigma_{abc} = 0$, {\it i.e.} the operator product $u_a u_b$ must vanish.  We have the following scenarios:
\begin{enumerate}[(a)]

\item $n_{\rm V} = 2$.  There is no $\bZ_3$-charged operator.

\item $n_{\rm V} = 4$.  There is a single pair of $\bZ_3$-charged operators.  Then \eqref{uuubar2} reads
\ie
\label{g=4Solution}
\sigma^2 = 1 + \xi^2 \, .
\fe

\item $n_{\rm V} = 6$, and there are two pairs of $\bZ_3$-charged operators with {\it different} $\xi_a$.  Because $\sigma_{abc}$ with mixed indices vanish, \eqref{uuubar2} becomes
\ie
\label{SigmaXi}
0 = 1 + \xi_1 \xi_2 \, , \quad \sigma_{111}^2 = 1 + \xi_1^2 \, , \quad \sigma_{222}^2 = 1 + \xi_2^2 \, .
\fe
We can use the residual ${\rm U}(1)^2$ basis freedom to make $\sigma_{aaa}$ real and non-negative.  Without loss of generality, 
\ie
\label{g=6Solution}
& \xi_1 = {\B - \sqrt{\B^2 + 4} \over 2} \, , 
\quad
\xi_2 = {\B + \sqrt{\B^2 + 4} \over 2} \, ,
\\
& \sigma_{111} = \sqrt{1+\xi_1^2} \, , 
\quad
\sigma_{222} = \sqrt{1+\xi_2^2} \, .
\fe

\item $n_{\rm V} = 6$, and there are two pairs of $\bZ_3$-charged operators with {\it the same} $\xi_a$.  It can be shown that the associativity of local operators admits a unique solution
\ie
\label{RuledOut}
\B = 2i \, , \quad \xi_1 = \xi_2 = i \, , \quad \sigma_{abc} = 0 \, .
\fe
This case will be ruled out momentarily.
\\
\end{enumerate}

To proceed, we examine the associativity of $\underline{o_{iA} o_{iB} v}$ detailed in Section~\ref{Sec:DefectAssociativity}.  The first condition \eqref{oovv}
\ie
\sum_{C} \kappa^i_{AC} \kappa^i_{BC} = \D_{AB} + \B \kappa^i_{AB}
\fe
implies that $\kappa^i_{AB}$ are $3 \times 3$ matrices with each eigenvalue taking one of two possible values
\ie
\label{kappaEigvals}
\text{each } {\rm eigval}(\kappa^i) = {\B \pm \sqrt{\B^2 + 4} \over 2} \, .
\fe
And it follows from the torus one-point modular invariance condition \eqref{T1xiv} that
\ie
{\rm tr}(\kappa^i) = - \zeta^{-1} \B \, , \quad \zeta = \frac{3 + \sqrt{13}}{2} \, .
\fe
We immediately see that \eqref{RuledOut} fails to satisfy this constraint, so (d) is ruled out.  In the following, we analyze the two inequivalent possibilities for the eigenvalues $\underline{---}$ and $\underline{+--}$ as labeled by the signs taken in \eqref{kappaEigvals}.  The $\underline{+++}$ and $\underline{++-}$ cases are equivalent to $\underline{---}$ and $\underline{+--}$ by the redefinition $v \to -v$.

\begin{enumerate}[I.]

\item $\underline{---}$ \quad
The torus one-point modular invariance condition \eqref{T1xiv} becomes
\ie
\label{Beta3}
3 \times {\B - \sqrt{\B^2 + 4} \over 2} = - \zeta^{-1} \B \quad\Rightarrow\quad \B = 3 \, .
\fe
As all eigenvalues of $\kappa^i_{AB}$ are the same, in any basis for $o_{iA}$,
\ie
\label{kappaBeta3}
\kappa^i_{AB} = - \zeta^{-1} \D_{AB} \, .
\fe
Besides the local operator algebra, the action of the $\rho$ TDL on the $\bZ_3$-charged operators is constrained as follows.  First, recall from \eqref{XiBarXi} that
\ie
\label{XiBarXi2}
\sum_c R_{ac} \overline{R}_{cb} = \D_{ab} \, .
\fe
Second, by the use of \eqref{xiuubarResult} and \eqref{kappaBeta3}, we can evaluate the left side of \eqref{xiuubarResult2},
\ie
\sum_{A,B} \C^i_{aA} \bar\C^i_{bB} \kappa^i_{AB} &= - \zeta^{-1} \sum_{A} \C^i_{aA} \bar\C^i_{bA} = - \D_{ab} \, ,
\fe
and then \eqref{xiuubarResult2} becomes
\ie
\label{B3}
- \D_{ab} = - \frac13 \left( \xi_{ab} + \sum_{c,d} R_{ad}\overline{R}_{bc} \xi_{dc} \right) \, .
\fe
Let us examine the scenarios (a)(b)(c).

\begin{enumerate}

\item
If $n_{\rm V} = 2$, then $\B = 3$ completely specifies the local operator algebra.

\item
If $n_{\rm V} = 4$, then \eqref{B3} becomes a scalar equation reading
\ie
-1 = -\frac13 \left( \xi + R \overline{R} \xi \right) = \frac23 \xi \, , \qquad \xi \equiv \xi_{11} = \xi_1 \, , \quad R \equiv R_{11} \, ,
\fe
which contradicts with the allowed $\xi$ values \eqref{xiSolved} given $\B = 3$.  Hence this case is ruled out.
  
\item
If $n_{\rm V} = 6$, then to be consistent with torus one-point modular invariance \eqref{Z3ModularInvariance} and the allowed $\xi_a$ values \eqref{xiSolved}, we set without loss of generality
\ie
\label{xiBeta3}
\xi_1 = - \zeta^{-1} \, , \quad \xi_2 = \zeta \, .
\fe
By \eqref{SigmaXi}, the non-vanishing three-point coefficients of $\bZ_3$-charged operators are
\ie
\label{sigma}
\sigma_{111} = \sqrt{1 + \zeta^{-2}} = \sqrt{13 - 3 \sqrt{13} \over 2} \, , \quad \sigma_{222} = \sqrt{1 + \zeta^2} = \sqrt{13 + 3 \sqrt{13} \over 2} \, .
\fe
We have thus completely specified the operator product algebra.  Together with \eqref{XiBarXi2} and \eqref{B3}, the action of the $\rho$ TDL on $\bZ_3$-charged local operators are restricted to be
\ie
\label{phasematrix}
\widehat\rho(u_a) =~ \x{.5}{$\rho_i$}{$u_a$} = \sum_b R_{ab} \bar u_b
\, ,
\quad
R = \theta \times
\begin{pmatrix}
0 & 1
\\
1 & 0
\end{pmatrix} \, ,
\quad
\theta \in \bC \, ,
\quad
|\theta| = 1 \, .
\fe
\end{enumerate}

\item $\underline{+--}$ \quad
The torus one-point modular invariance condition \eqref{T1xiv} becomes
\ie
{3 \B - \sqrt{\B^2 + 4} \over 2} = - \zeta^{-1} \B \, , \quad\Rightarrow\quad \B = {1\over\sqrt3} \, .
\fe
The values of $\xi_{ab}$ and $\sigma_{abc}$ are fixed by $\B$ through \eqref{xiSolved}, \eqref{g=4Solution}, and \eqref{g=6Solution}.  The bootstrap analysis of this possibility is more complicated than the $\underline{---}$ case, so we leave it for future work.  However, some hints pointing towards the existence of a defect TFT of case II(b) with $n_{\rm V} = 4$, and arguments for the non-existence in cases II(a) with $n_{\rm V} = 2$ and II(c) with $n_{\rm V} = 6$ can be found in Section~\ref{Sec:NIM}.

\end{enumerate}

In the next section, we complete the construction of a TFT of case I(c) with $n_{\rm V} = 6$ and $\B = 3$.  The reader interested in boundary conditions can safely proceed to Section~\ref{Sec:NIM}.

\subsection{Topological field theory with six vacua}
\label{Sec:TFT6}

We now construct the rest of the defect TFT data in case I(c) with $n_{\rm V} = 6$ and $\B = 3$, and solve {\bf all} the consistency conditions outlined in Section~\ref{Sec:Crossing}.

It turns out that a good point of attack is the associativity of $\underline{o_{iA} o_{iB} v}$.  The condition \eqref{oovu} in the basis diagonalizing $\xi_{ab}$ \eqref{tau} reads
\ie
\label{Eigen}
\sum_{C} \kappa^i_{AC} \lambda^i_{BC;a} = \xi_a \lambda^i_{AB;a} \, ,
\fe
which implies that for fixed $A$ and $a$, $\lambda^i_{AB;a}$ must be an eigenvector of $\kappa^i$ with eigenvalue $\xi_a$; otherwise $\lambda^i_{BC;a}$ vanishes.  But because $\kappa^i$ does not have $\zeta$ as an eigenvalue, it follows that
\ie
\lambda^i_{AB;2} = 0 \, .
\fe
Note that the vanishing of $\lambda^i_{AB;2}$ is consistent with \eqref{oouu}.  

By considering the vanishing $\lambda^i_{AB;2}$, we can determine the $\rho$ action, which we found to be parameterized by $\theta \in \bC$ in \eqref{phasematrix}.  The nontrivial part of \eqref{xiuuResult} with $f = 2$ becomes
\ie
0 = \zeta \sigma_{111} \bar\theta - \theta^2 \sigma_{222} \, ,
\fe
which by the use of \eqref{sigma} leads to
\ie
|\theta| = \zeta {\sigma_{111} \over \sigma_{222}} = \zeta \sqrt{1 + \zeta^{-2} \over 1 + \zeta^2} = 1 \, , \quad \theta^3 = 1 \, .
\fe
Up to the relabeling of $\rho_i$,
\ie
\label{Theta1}
\theta = 1 \, .
\fe

For the non-vanishing $\lambda^i_{AB;1}$, it is convenient to define a normalized 
\ie
\hat\lambda^i_{AB} \equiv 
{\lambda^i_{AB;1} \over \sigma_{111}} \, ,
\fe
and write the associativity of $\underline{o_{iA} o_{iB} u_a}$ \eqref{oouu} and the modular invariance condition \eqref{TrLambda} in matrix notation as (recall that $\hat\lambda^i$ is a symmetric matrix)
\ie
\label{LambdaEquations}
\hat\lambda^i \bar{\hat\lambda}^i = 1 \, , \quad \hat\lambda^i \hat\lambda^i = \bar{\hat\lambda}^i \, , \quad {\rm Tr} \, \lambda^i = 0 \, .
\fe
The first equation says that $\hat\lambda^i$ is unitary, and combined with the second equation implies that $(\hat\lambda^i)^3 = 1$.  The third equation then tells us that $\hat\lambda^i$ has eigenvalues
\ie
{\rm eigvals}(\hat\lambda^i) = \{1, \, \omega, \, \omega^2\} \, . 
\fe

We now prove that $\hat\lambda$ (suppressing superscript $i$) must be diagonalizable by an ${\rm O}(3)$ matrix.  For convenience define $\Omega = \text{diag}(1, \omega, \omega^2)$.  Because $\hat\lambda$ is unitary, it can always be diagonalized by a unitary matrix $Z$, {\it i.e.} $\hat\lambda = Z^\dag \Omega Z$.  For $\hat\lambda$ to be symmetric, we must have
\ie
\label{ZWZ}
Z^\dagger W Z = Z^T W \bar Z \quad\Rightarrow\quad (Z Z^T) W \overline{(Z Z^T)} = W.
\fe
Let us define $A=Z Z^T$. The $(1,1)$-component of the matrix equation \eqref{ZWZ} reads
\ie
|A_{11}|^2+\omega |A_{12}|^2 + \omega^2 |A_{13}|^2=1 \, ,
\fe
where we used the fact that $A$ is symmetric. Now for the above equation to have a solution, we must have $|A_{12}|^2=|A_{13}|^2$, since otherwise the imaginary part cannot match. Let us call this value $x \equiv |A_{12}|^2=|A_{13}|^2$. Then $|A_{11}|^2=1+x$. Proceeding similarly, we arrive at the following matrix
\ie
A=\begin{pmatrix}
   e^{i a_{11}}\sqrt{1+x} & e^{i a_{12}}\sqrt{x} & e^{i a_{13}}\sqrt{x}\\
   e^{i a_{12}}\sqrt{x} & e^{i a_{22}}\sqrt{1+x} & e^{i a_{23}}\sqrt{x}\\
   e^{i a_{13}}\sqrt{x} & e^{i a_{23}}\sqrt{x} & e^{i a_{33}}\sqrt{1+x}\\
  \end{pmatrix},
\fe
where $a_{ij}$ are arbitrary phases. Finally, $A$ must be unitary since
\ie
A A^\dagger = (Z Z^T)(Z Z^T)^\dagger = Z (Z^T \bar Z) Z^\dagger = Z Z^\dagger = 1 \, ,
\fe
which means that
\ie
(AA^\dagger)_{11}=1+3x=1 \, .
\fe
Hence $x=0$ and we end up with
\ie
A=\begin{pmatrix}
   e^{i a_{11}} &  & \\
    & e^{i a_{22}} & \\
    &  & e^{i a_{33}}\\
  \end{pmatrix} \, .
\fe
The ${\rm O}(3)$ matrix of interest is given by\footnote{The fact that $O$ is an ${\rm O}(3)$  matrix follows from $O O^\dag = O O^T = 1$.  The first equality implies that $O^\dag = O^T$, or equivalently that $O$ is real, and the second equality is orthogonality.
}
\ie
O \equiv \sqrt{A^{-1}} Z \, .
\fe
We can therefore use the ${\rm O}(3)$ freedom to set 
\ie
\label{lambda}
\hat\lambda^i_{AB;1} = \omega^{A-1-i} \D_{AB} \quad\Rightarrow\quad
\lambda^i_{AB;1} = \omega^{A-1-i} \sigma_{111} \D_{AB} = \omega^{A-1-i} \sqrt{1+\zeta^{-2}} \, \D_{AB} \, .
\fe

Let us summarize the solution we found so far into
\ie
& v \times v = 1 + 3 v \, , \quad u_1 \times \bar u_1 = 1 - \zeta^{-1} v \, , \quad u_2 \times \bar u_2 = 1 + \zeta v \, ,
\\
\\
& u_1 \times \bar u_2 = 0 \, , \quad u_1 \times u_1 = \sqrt{1 + \zeta^{-2}} \, \bar u_1 \, , \quad u_2 \times u_2 = \sqrt{1 + \zeta^2} \, \bar u_2 \, ,
\\
\\
& \defect{.5}{0}{$\rho_i$}{$o_{iA}$}{$o_{iB}$} = 
\begin{cases}
\displaystyle
1 - \zeta^{-1} v + \sqrt{1 + \zeta^{-2}} \left( \omega^{i+1-A} u_1 + \omega^{A-1-i} \bar u_1 \right) & A = B \, ,
\\
0 & A \neq B \, .
\end{cases}
\fe
We proceed to solve the more general crossing symmetry involving four $\rho_i$ defect operators.  Some analytic progress is made in Appendix~\ref{App:Defect}, such as deriving a selection rule \eqref{SelectionRule}, but eventually we resort to computer numerics to find a solution.\footnote{Up to this point in the main text, no assumption about reflection-positivity was needed.  However, both Appendix~\ref{App:Defect} and the computer numerics assume reflection-positivity.
}
Up to operator relabeling and sign redefinitions, the solution appears to be unique.  The non-vanishing defect three-point coefficients are (vacuum expectation values are implicitly taken)
\ie
\label{3PtSolution}
\tri{.5}{0}{$o_{iA}$}{$o_{iA}$}{$o_{iA}$} \ &= \ \sqrt{\frac{17 \sqrt{13}-52}{3}} \, ,
\\
\tri{.5}{0}{$o_{i+1,A+1}$}{$o_{iA}$}{$o_{iA}$} \ &= \ - \frac{1}{2} \sqrt{\frac{1}{3} \left(-13+11 \sqrt{13}+\sqrt{78 \left(7 \sqrt{13}-23\right)}\right)} \, ,
\\
\tri{.5}{0}{$o_{i-1,A-1}$}{$o_{iA}$}{$o_{iA}$} \ &= \ \frac{1}{2} \sqrt{\frac{1}{3} \left(-13+11 \sqrt{13}-\sqrt{78 \left(7 \sqrt{13}-23\right)}\right)} \, ,
\\
\tri{.5}{0}{$o_{iA}$}{$o_{i+1,A+1}$}{$o_{i-1,A-1}$} \ &= \
\tri{.5}{0}{$o_{iA}$}{$o_{i-1,A-1}$}{$o_{i+1,A+1}$} ~=~-\sqrt{\frac{13+7 \sqrt{13}}{6}} \, .
\fe
Note an interesting ``superselection'' rule:  If we define three ``sectors'' labeled by $i - A \mod 3$, then all non-vanishing three-point coefficients are those that involve defect operators in a single sector.

Finally, we can solve the full set of modular invariance constraints, which are linear in the lassos.  We find a solution where some of the lassos are given by (vacuum expectation values are implicitly taken)
\ie
& 
\bell{1}{$o_{iA}$}{$\rho_i$}{$v$} = \varepsilon^i_A
= - \sqrt{\sqrt{13} \over 3} \, ,
\\
& 
\bell{1}{$o_{iA}$}{$\rho_i$}{$u_1$} = \C^i_{1A}
= \sqrt{\zeta \over 3} \, \omega^{A-1-i} e^{-i\phi_\C} \, ,
\quad
\phi_\C = {\pi\over4} + {1\over2} \arccos{1\over3+\sqrt{13}} \, ,
\\
& 
\bell{1}{$o_{iA}$}{$\rho_i$}{$u_2$} = \C^i_{2A}
= - \sqrt{\zeta \over 3} \, \omega^{1-A-i} \, ,
\\
&
\lassoX{1}{$o_{iB}$}{$\rho_i$}{$\rho_i$}{$o_{iA}$} = 
\begin{cases}
\displaystyle {-1 + 5\sqrt{13} \over 18} & A = B \, ,
\\
\\
\displaystyle {-7 - \sqrt{13} \over 18} & A \neq B \, ,
\end{cases}
\\
&
\lassoX{1}{$o_{iB}$}{$\rho_{i\pm1}$}{$\rho_{i\mp1}$}{$o_{iA}$} = 
\begin{cases}
\displaystyle \frac{17+5\sqrt{13}}{18} & A = B \, ,
\\
\\
\displaystyle \frac{-7-\sqrt{13}}{18} & A \neq B \, ,
\end{cases}
\\
&
\lassoX{1}{$o_{iB}$}{$\rho_{i\pm1}$}{$\rho_{i\pm1}$}{$o_{iA}$} = 
\begin{cases}
\displaystyle \frac{5+2 \sqrt{13}\pm\sqrt{15+6 \sqrt{13}}}{18} & A = B \, ,
\\
\\
\displaystyle \frac{-11+\sqrt{13}\pm\sqrt{6 \left(61+19 \sqrt{13}\right)}}{36} & A \neq B \, ,
\end{cases}
\\
&
\lassoX{1}{$o_{iB}$}{$\rho_i$}{$\rho_{i\pm1}$}{$o_{iA}$} = 
\begin{cases}
\displaystyle \frac{5+2 \sqrt{13}\pm\sqrt{15+6 \sqrt{13}}}{18} & A = B \, ,
\\
\\
\displaystyle \frac{7+\sqrt{13}\mp\sqrt{798+222 \sqrt{13}}}{36} & A \neq B \, ,
\end{cases}
\fe
and the rest are related to the above via \eqref{LassoRelations}.  

We have thus completed the construction of a defect topological field theory whose defining data solve {\bf all} the consistency conditions outlined in Section~\ref{Sec:Crossing}.

\section{Boundary conditions and NIM-reps}
\label{Sec:NIM}

We can extend our topological field theory (TFT) further by considering boundaries.  Given the bijection between $\cC$-symmetric TFTs and $\cC$-module categories argued by Thorngren and Wang \cite{Thorngren:2019iar}, the fact that the Haagerup $\cH_3$ fusion category has exactly three indecomposable module categories \cite{Grossman_2012}, with two, four, and six simple objects, is strongly suggestive of a connection to the three ``minimal'' possible TFTs (with $n_{\rm P} = 15$ point-like operators) in Table~\ref{Tab:Possibilities}, with two, four, and six vacua.  However, except in special cases, it is not known how to extract the axiomatic defect TFT data from the module category.  In this section, without assuming any prior knowledge of module categories, we use the bootstrap results on the axiomatic defect TFT data to construct boundaries and examine their fusion with TDLs.  This section can be read independently of Section~\ref{Sec:TFT6}.

We first review some nontrivial results in open/closed TFTs.  The admissible boundary conditions of a TFT are direct sums of a set of elementary boundary conditions $B_a$, which are related to the so-called Cardy states $\nu_a$ by folding the boundary into a circle and invoking the state-operator map,
\ie
\begin{gathered}
\begin{tikzpicture}[scale=.5]
\draw [line] (0,-2) node[below] {$B_a$} -- (0,0) node[left=3pt] {TFT} -- (0,2);
\fill [fill={rgb,255:red,200; green,200; blue,200},opacity=0.5] (0,-2) rectangle (1,2);
\end{tikzpicture}
\end{gathered}
\quad \rightarrow \quad
\begin{gathered}
\begin{tikzpicture}[scale=.5]
\draw [line] (0,0)  arc (90:270:0.5 and 1.5);
\draw [line,dashed] (0,-3) arc (-90:90:0.5 and 1.5);
\draw [bg] (-3,0) -- (3,0);
\draw [bg] (-3,-3) -- (3,-3);
\fill [fill={rgb,255:red,200; green,200; blue,200},opacity=0.5] (0,-3) rectangle (3,0);
\fill [fill={rgb,255:red,200; green,200; blue,200},opacity=0.5] (0,0)  arc (90:270:0.5 and 1.5);
\draw (-.5,-1.5) node[left] {TFT};
\draw (.5,-1.5) node[right] {$|\nu_a\rangle$};
\end{tikzpicture}
\end{gathered}
\quad \mapsto \quad
\begin{gathered}
\begin{tikzpicture}[scale=1]
\draw [bg] (0,0) -- (1,2) -- (4,2) -- (3,0) -- (0,0);
\draw [line] (2,1) \dt{below}{0}{$\nu_a$};
\end{tikzpicture}
\end{gathered}
\fe
By solving the consistency conditions of open/closed TFT, Moore and Segal \cite{Moore:2006dw} established an explicit formula for the Cardy states $\n_a$ in terms of the projectors $\pi_a$ introduced in Section~\ref{Sec:Frobenius},
\ie
\label{MooreSegal}
\n_a = \frac{\pi_a}{\sqrt{\vev{\pi_a}}} \, .
\fe
In particular, the number of Cardy states is the same as the number of vacua $n_{\rm V}$.

Let $\{ \cL_i \mid i = 1, \dotsc, r \}$ be the set of simple TDLs, $N_{ij}^k$ the fusion coefficients, and $\{ \n_a \mid a = 1, \dotsc, n_{\rm V} \}$ the set of Cardy states.  The fusion of any TDL with an admissible boundary must give another admissible boundary.  Therefore, the action of TDLs on the Cardy states must furnish a non-negative integral matrix representation (NIM-rep):  a set of $n_{\rm V} \times n_{\rm V}$ non-negative integral matrices $(\cN_i)_a^{~b}$, one for each line $i$, such that
\ie
\label{NIM}
\sum_c (\cN_i)_a^{~c} (\cN_j)_c^{~b} = N_{ij}^k \, (\cN_k)_a^{~b} \, .
\fe

Given that we have narrowed down the full local operator algebra to a few possibilities in Section~\ref{Sec:LocalAlgebra}, it is straightforward to compute the projector basis and examine the action of the TDLs on the projectors.  To condense the discussion, we present formulae that apply to the entire family of local operator algebras, parameterized by $\B$, that solve the associativity of local operators.  Of course, we have seen that the associativity of $\underline{o_{iA} o_{iB} v}$ requires $\B = 3$ or $\B = \frac{1}{\sqrt3}$.

\begin{enumerate}[(a)]

\item
Consider $n_{\rm V} = 2$.  The projector basis for
\ie
v \times v = 1 + \beta v
\fe
is given by
\ie
\pi_1 = \frac{\zeta - v}{\sqrt{4+\beta^2}} \, , \quad \pi_2 = \frac{\zeta^{-1} + v}{\sqrt{4+\beta^2}} \, .
\fe
According to \eqref{MooreSegal}, the Cardy states are
\ie
\n_1 = \frac{\sqrt[4]{4+\beta^2}}{\sqrt\zeta} \, \pi_1 \, , \quad 
\n_2 = \sqrt[4]{4+\beta^2} \sqrt\zeta \, \pi_2 \, .
\fe

    \begin{enumerate}[I.]
    
    \item
    When $\beta=3$, they furnish a NIM-rep 
    \ie
    \cN_\A = \begin{pmatrix}
    	1
    	\\
    	& 1
    \end{pmatrix} \, , 
    \quad
    \cN_\rho = \begin{pmatrix}
    	3 & 1
    	\\
    	1 &
    \end{pmatrix} \, .
    \fe
    
    \item
    When $\beta=\frac{1}{\sqrt{3}}$, the representation
    \ie
    \cN_\A = \begin{pmatrix}
    	1
    	\\
    	& 1
    \end{pmatrix} \, , 
    \quad
    \cN_\rho = \begin{pmatrix}
    	2 & \sqrt{3}
    	\\
    	\sqrt{3} & 1
    \end{pmatrix}
    \fe
    is not NIM.
    
    \end{enumerate}

\item
For $n_{\rm V}=4$, the local operator algebra is given by
\ie
& v \times v = 1 + \beta v \, , \quad u \times \bar u = 1 +\xi v \, ,\quad u \times u = \sqrt{1 + \xi^2} \, \bar u \,, \quad \xi=\frac{\beta\pm\sqrt{\beta^2+4}}{2} \, .
\fe
There are two possible choices for $\xi$, and we can construct the projector basis
\ie
\pi_a &= \begin{cases}
	\displaystyle
	\frac{\epsilon\,\xi^{-1} +\epsilon\, v + \sqrt[4]{4+\beta^2} \, \sqrt{\epsilon\,\xi^{-1}} \, (\omega^{a-1} u + \omega^{1-a} \bar u)}{3\sqrt{4+\beta^2}} \, , & a = 1, 2, 3 \, ,
	\\
	\displaystyle
	\frac{\epsilon\,\xi -\epsilon\, v}{\sqrt{4+\beta^2}} \, , & a = 4\, ,
\end{cases}
\fe
where $\epsilon \equiv {\rm sign}(\xi)$.  The Cardy states \eqref{MooreSegal} are then
\ie
\n_a = \begin{cases}
	\displaystyle\left(\frac{\epsilon\,\xi^{-1}}{3\sqrt{4+\beta^2}}\right)^{-\frac{1}{2}}\pi_a \, , & a = 1, 2, 3 \, ,
	\\
	\displaystyle\left(\frac{\epsilon\,\xi}{\sqrt{4+\beta^2}}\right)^{-\frac{1}{2}} \pi_a \, , & a = 4 \, .
\end{cases}
\fe
Whether they furnish a NIM-rep depends on how the $\rho$ TDL acts, that is, on $R$.  

    \begin{enumerate}[I.]
    
    \item 
    When $\beta=\frac{1}{\sqrt{3}}$, we find that for $\xi$ taking either value, that is $\epsilon = \pm$, there is exactly one value of $R$ that gives rise to a NIM-rep:
    \ie
    \epsilon = + \, , \quad
    R=1 &:\, \quad \cN_\A = \begin{pmatrix}
    	&~1~
    	\\
    	&&~1~
    	\\
    	~1~&& &
    	\\
    	&&&~1~
    \end{pmatrix} \, , 
    \quad
    \cN_\rho = \begin{pmatrix}
    	~1~ &  &  & ~1~ \\
    	 & & ~1~ & ~1~  \\
    	 & ~1~ &  & ~1~ \\
    	~1~ &~1~  &~1~  &~2~
    \end{pmatrix} \, ,
    \\
    \epsilon = - \, , \quad
    R=-1 & :\, \quad \cN_\A = \begin{pmatrix}
    	&~1~
    	\\
    	&&~1~
    	\\
    	~1~&& &
    	\\
    	&&&~1~
    \end{pmatrix} \, , 
    \quad
    \cN_\rho = \begin{pmatrix}
    	 & ~1~ & ~1~ & ~1~ \\
    	~1~ & ~1~ & & ~1~  \\
    	~1~ &  & ~1~ & ~1~ \\
    	~1~ & ~1~ & ~1~ & ~1~
    \end{pmatrix} \, .
    \fe
    
    \item 
    When $\beta=3$, we instead have
    \ie
    \hspace{-.1in}
    \epsilon = +
    &: ~ \cN_\rho = \begin{pmatrix}
    	~\frac{2\cos\phi}{3}~ & ~-\frac{\cos\phi}{3}+\frac{\sin\phi}{\sqrt{3}}~ & ~-\frac{\cos\phi}{3}-\frac{\sin\phi}{\sqrt{3}}~ & ~\frac{1}{\sqrt{3}}~ \\
    	~-\frac{\cos\phi}{3}+\frac{\sin\phi}{\sqrt{3}}~ & ~-\frac{\cos\phi}{3}-\frac{\sin\phi}{\sqrt{3}}~ & ~\frac{2\cos\phi}{3}~ & ~\frac{1}{\sqrt{3}}~  \\
    	~-\frac{\cos\phi}{3}-\frac{\sin\phi}{\sqrt{3}}~ & ~\frac{2\cos\phi}{3}~ & ~-\frac{\cos\phi}{3}+\frac{\sin\phi}{\sqrt{3}}~ & ~\frac{1}{\sqrt{3}}~ \\
    	~\frac{1}{\sqrt{3}}~ & ~\frac{1}{\sqrt{3}}~ & ~\frac{1}{\sqrt{3}}~ & ~3~
    	\end{pmatrix} \, ,
    \\
    \hspace{-.1in} \epsilon = -
    &: ~ \cN_\rho = \begin{pmatrix}
    	~1+\frac{2\cos\phi}{3}~ & ~1-\frac{\cos\phi}{3}+\frac{\sin\phi}{\sqrt{3}}~ & ~1-\frac{\cos\phi}{3}-\frac{\sin\phi}{\sqrt{3}}~ & ~\frac{1}{\sqrt{3}}~ \\
    	~1-\frac{\cos\phi}{3}+\frac{\sin\phi}{\sqrt{3}}~ & ~1-\frac{\cos\phi}{3}-\frac{\sin\phi}{\sqrt{3}}~ & ~1+\frac{2\cos\phi}{3}~ & ~\frac{1}{\sqrt{3}}~  \\
    	~1-\frac{\cos\phi}{3}-\frac{\sin\phi}{\sqrt{3}}~ & ~1+\frac{2\cos\phi}{3}~ & ~1-\frac{\cos\phi}{3}+\frac{\sin\phi}{\sqrt{3}}~ & ~\frac{1}{\sqrt{3}}~ \\
    	~\frac{1}{\sqrt{3}}~ & ~\frac{1}{\sqrt{3}}~ & ~\frac{1}{\sqrt{3}}~ & ~0~
    	\end{pmatrix} \, ,
    \fe
    where $R = e^{i\phi}$.  As we can see the representation is not NIM.
    
    \end{enumerate}

\item
Finally, consider $n_{\rm V} = 6$.  The local operator algebra is
\ie
& v \times v = 1 + \beta v \, , \quad u_1 \times \bar u_1 = 1 +\xi_1 v \, , \quad u_2 \times \bar u_2 = 1 + \xi_2 v \, ,
\\
& u_1 \times \bar u_2 = 0 \, , \quad u_1 \times u_1 = \sqrt{1 + \xi_1^2} \, \bar u_1 \, , \quad u_2 \times u_2 = \sqrt{1 + \xi_2^2} \, \bar u_2 \, , 
\\
& \xi_{1,2}=\frac{\beta\mp\sqrt{\beta^2+4}}{2} \, ,
\fe
and the projector basis is given by
\ie
\hspace{-.5in}
\pi_a &= \begin{cases}
	\displaystyle
	\frac{\xi_2 - v + \sqrt[4]{4+\beta^2} \, \sqrt{\xi_2} \, (\omega^{a-1} u_1 + \omega^{1-a} \bar u_1)}{3\sqrt{4+\beta^2}} \, , & a = 1, 2, 3 \, ,
	\\
	\displaystyle
	\frac{-\xi_1 + v + \sqrt[4]{4+\beta^2} \, \sqrt{-\xi_1} \, (\omega^{a-1} u_2 + \omega^{1-a} \bar u_2)}{3\sqrt{4+\beta^2}} \, , & a = 4, 5, 6 \, .
\end{cases}
\fe
The Cardy states \eqref{MooreSegal} are
\ie
\n_a = \begin{cases}
	\displaystyle\left(\frac{\xi_2}{3\sqrt{4+\beta^2}}\right)^{-\frac{1}{2}}\pi_a \, , & a = 1, 2, 3 \, ,
	\\
	\displaystyle\left(\frac{-\xi_1}{3\sqrt{4+\beta^2}}\right)^{-\frac{1}{2}} \pi_a \, , & a = 4, 5, 6 \, .
\end{cases}
\fe
Clearly, the triples $(\nu_1, \nu_2, \nu_3)$ and $(\nu_4, \nu_5, \nu_6)$ each transforms as a three-dimensional permutation representation under $\bZ_3$.

    \begin{enumerate}[I.]
    
    \item 
    When $\beta=3$, with the $\rho$ TDL action $R_{ab}$ given by \eqref{phasematrix} and \eqref{Theta1}, one can check that the Cardy states furnish a NIM-rep
    \ie
    \cN_\A = \begin{pmatrix}
    	&~1~
    	\\
    	&&~1~
    	\\
    	~1~
    	\\
    	&&& &~1~
    	\\
    	&&& &&~1~
    	\\
    	&&& ~1~
    \end{pmatrix} \, , 
    \quad
    \cN_\rho = \begin{pmatrix}
    	~1~ & ~1~ & ~1~ & ~1~ &  &  \\
    	~1~ & ~1~ & ~1~ &  & ~1~ &  \\
    	~1~ & ~1~ & ~1~ &  &  & ~1~ \\
    	~1~ &  &  &  &  &  \\
    	& ~1~ &  &  &  &  \\
    	&  & ~1~ &  &  &  \\
    \end{pmatrix} \, .
    \fe
    
    \item
    For $\beta=\frac{1}{\sqrt{3}}$, without assuming anything about the matrix $R_{ab}=x_{ab}+i y_{ab}$, we get the following representation for $\rho$ action:
    \ie
    \hspace{-.1in} \cN_\rho = \begin{pmatrix}
    	A&B\\C&D
    \end{pmatrix},
    ~~
    B=\begin{pmatrix}
    	~\frac{1}{\sqrt{3}}+\frac{2x_{12}}{3}~& ~-\frac{x_{12}}{3}+\frac{1+y_{12}}{\sqrt{3}}~ & ~-\frac{x_{12}}{3}+\frac{1-y_{12}}{\sqrt{3}}~\\
    	~-\frac{x_{12}}{3}+\frac{1+y_{12}}{\sqrt{3}}~ & ~-\frac{x_{12}}{3}+\frac{1-y_{12}}{\sqrt{3}}~&~\frac{1}{\sqrt{3}}+\frac{2x_{12}}{3}~ \\
    	 ~-\frac{x_{12}}{3}+\frac{1-y_{12}}{\sqrt{3}}~&~\frac{1}{\sqrt{3}}+\frac{2x_{12}}{3}~&~-\frac{x_{12}}{3}+\frac{1+y_{12}}{\sqrt{3}}~  
    \end{pmatrix} \, ,
    \fe
    and $A, \, C, \, D$ are other $3 \times 3$ matrices whose explicit form we do not need.  Suppose $B$ is NIM.  Because $B_{12}-B_{13}=\frac{2y_{12}}{\sqrt{3}}$, it follows that $y_{12}$ must be a multiple of $\frac{\sqrt{3}}{2}$, and we can write $y_{12}=\frac{n\sqrt{3}}{2}$ with $n\in\bZ$. But then $B_{11}+2B_{12}=\sqrt{3}+n$. Hence no NIM-rep exists.
    
    \end{enumerate}

\end{enumerate}

The results of the above analysis are summarized in Table~\ref{Tab:Existence}.  The defect TFT constructed in Section~\ref{Sec:TFT6} passed the NIM-rep test, and notably the NIM-rep requirement in case II(b) allowed us to determine the action of the $\rho$ TDL on the $\bZ_3$-charged operators.

\begin{table}[H]
\centering
\begin{tabular}{|c||c|c|c|}
\hline
& (a) $n_{\rm V} = 2$ & (b) $n_{\rm V} = 4$ & (c) $n_{\rm V} = 6$
\\\hline\hline
I. ~ $\B = 3$ & $\circ$ & $\times$ & \cellcolor{Blue} $\circ$
\\
II. ~ $\B = \frac{1}{\sqrt3}$ & $\times$ & $\circ$ & $\times$
\\\hline
\end{tabular}
\caption{Existence of (1+1)$d$ topological field theories realizing the Haagerup $\cH_3$ fusion category from analyzing the fusion of topological defect lines with the admissible boundary conditions.  We restrict to theories with exactly two $\bZ_3$-neutral vacua $1$ and $v$.  Here $n_{\rm V}$ denotes the total number of vacua, and $\B$ is the coefficient in the fusion rule $v \times v = 1 + \B v$.  The $\circ$ marks the cases that pass the NIM-rep condition, and the $\times$ marks those ruled out.  The theory constructed in Section~\ref{Sec:TFT6} is highlighted.
}
\label{Tab:Existence}
\end{table}

\section{Realization of Haagerup $\cH_1$ and $\cH_2$ via gauging}
\label{Sec:Gauging}

Given (a (1+1)$d$ quantum field theory with) a finite symmetry group $G$ that contains a non-anomalous subgroup $H$, gauging $H < G$ gives rise to (a quantum field theory with) a fusion category symmetry $F'$ that contains a ${\rm Rep}(H)$ sub-category.  This process can be reversed by gauging ${\rm Rep}(H) < F'$.  In this sense, the pairs $(G, H)$ and $(F, \, {\rm Rep}(H))$ are dual to each other.  A generalization of the above statement is the following: given a fusion category $\cC$ that contains an algebra object (a non-simple topological defect line satisfying certain conditions) $A$, gauging $A < \cC$ gives rise to a fusion category $\cC' = {\rm Bimod}_\cC(A, A)$ (category of $(A, A)$ bimodules within $\cC$) that contains a dual algebra object $A'$, and this process can be reversed by gauging $A' < \cC'$.  Thus, the pairs $(\cC, A)$ and $(\cC', A')$ are dual to each other.\footnote{Gauging by different algebra objects $A_1$ and $A_2$ with the same module category ${\rm Mod}_\cC(A_1) = {\rm Mod}_\cC(A_2)$ gives rise to the same gauged theory, so $A_1$ and $A_2$ are equivalent in the context of gauging.  The duality pairing of $(\cC, A)$ and $(\cC', A')$ is up to this equivalence.
}
The reader is referred to \cite{Bhardwaj:2017xup} for a much more refined discussion, and to \cite{Fuchs:2002cm,Frohlich:2009gb,Carqueville:2012dk} for the original idea of generalized gauging.

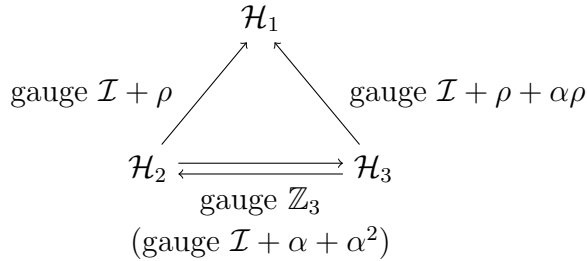
\begin{figure}[H]
\begin{center}
\begin{tikzpicture}[scale=0.5]
\draw (3,4) node (H1) {$\cH_1$};
\draw (0,0) node (H2) {$\cH_2$};
\draw (6,0) node (H3) {$\cH_3$};
\draw (-1.5,2) node {gauge $\cI+\rho$};
\draw (8.5,2) node {gauge $\cI+\rho+\alpha\rho$};
\draw (3,0) node[below=3pt] {\parbox{6cm}{\centering gauge $\bZ_3$ \\ (gauge $\cI + \A + \A^2$)}};
\draw [->] (H2.60) -- (H1.240);
\draw [->] (H3.120) -- (H1.300);
\draw [->] (H2.10) -- (H3.170);
\draw [<-] (H2.350) -- (H3.190);
\end{tikzpicture}
\end{center}
\caption{Gauging relations among theories realizing the three Haagerup fusion categories.
}
\label{Fig:Triangle}
\end{figure}

The relations among the Haagerup $\cH_1, \, \cH_2, \, \cH_3$ fusion categories can be understood this way.  Up to automorphism, there are two nontrivial algebra objects in $\cH_2$, one corresponding to the non-anomalous $\bZ_3$ symmetry $\cI + \A + \A^2$, and the other to $\cI + \rho$.  There are also two nontrivial algebra objects in $\cH_3$, one again corresponding to the non-anomalous $\bZ_3$ symmetry $\cI + \A + \A^2$, and the other to $\cI + \rho + \A\rho$.  Gauging the $\bZ_3$ symmetry exchanges $\cH_2$ and $\cH_3$, and gauging the other nontrivial algebra object in either $\cH_2$ or $\cH_3$ gives $\cH_1$.  These relations are summarized in Figure~\ref{Fig:Triangle}.

Thus, to construct topological field theories realizing the Haagerup $\cH_1$ or $\cH_2$ fusion category, one can simply take a topological field theory realizing $\cH_3$, such as the one we constructed in Section~\ref{Sec:TFT6}, and gauge \,$\cI + \rho + \A\rho$\, or \,$\cI + \A + \A^2$ (the $\bZ_3$ symmetry), respectively.  A discussion on the gauging of algebra objects in (1+1)$d$ topological field theory can be found in \cite{Bhardwaj:2017xup}.  In particular, gauging the theory we constructed, which has $n_{\rm V} = 6$ vacua and realizes the Haagerup $\cH_3$ fusion category, by $\bZ_3$ gives rise to a theory that has $n_{\rm V} = 2$ vacua and realizes the Haagerup $\cH_2$ fusion category.

\section{Prospective questions}
\label{Sec:Remarks}

\begin{itemize}

\item  What is the full axiomatic data, when boundaries are included, of the defect topological field theory that we constructed in Section~\ref{Sec:TFT6}?

\item  Is there an explanation for the ``superselection'' rule noted below \eqref{3PtSolution}?

\item  The construction of topological field theories realizing cases I(a), with $n_{\rm V} = 2$ vacua and $\B = \frac{1}{\sqrt3}$, and II(b), with $n_{\rm V} = 4$ vacua and $\B = \frac{1}{\sqrt3}$, is left for future work.  For these cases, we showed in Section~\ref{Sec:NIM} that the Cardy states obtained from bootstrap furnish non-negative integer matrix representations under fusion with topological defect lines.

\item  Is there a conformal field theory realizing Haagerup or its quantum double?  Despite nontrivial positive evidence from the work of Evans and Gannon \cite{Evans:2010yr}, and recent attempts by Wolf \cite{Wolf:2021kkq}, the question remains open.  The defect modular bootstrap approach of \cite{Lin:2019kpn,Lin:2021udi} may put universal constraints on such conformal field theories.

\item  Is Haagerup truly {\it exotic} (whatever exotic means)?  Evans and Gannon \cite{Evans:2010yr} suggested not, as it sits inside a hypothetically infinite family of Haagerup-Izumi subfactors/fusion categories \cite{izumi2001structure}.  The transparent $F$-symbols for some higher members of this family have been recently computed by the present authors \cite{Huang:2020lox}, and may allow for the construction of the corresponding defect topological field theories.

\item Finally, the broader questions {\bf Q1}, {\bf Q2}, and {\bf Q4} of Section~\ref{Sec:Intro} motivating this work remain open.

\end{itemize}

\section*{Acknowledgements}

We thank Yuji Tachikawa for suggesting this interesting project, for helpful advice, and for comments on the first draft.  We also benefited from enlightening discussions with Chi-Ming Chang, Kantaro Ohmori, Sahand Seifnashri, Shu-Heng Shao, and Yifan Wang.  This material is based upon work supported by the U.S. Department of Energy, Office of Science, Office of High Energy Physics, under Award Number DE-SC0011632.  YL is supported by the Simons Collaboration Grant on the Non-Perturbative Bootstrap and by the Sherman Fairchild Foundation.

\appendix

\section{The $F$-symbols for the Haagerup $\cH_3$ fusion category}
\label{App:CrossingKernel}

This appendix presents the $F$-symbols for the transparent Haagerup $\cH_3$ fusion category found in \cite{Huang:2020lox}.  We first present the unitary gauge, and then transit to a slightly more convenient gauge for this note.

Let $I = \{ \cI, \, \A, \, \A^2 \}$ be the set of invertible objects, $N = \{ \rho, \, \A\rho, \, \A^2\rho\}$ be the set of non-invertible simple objects of the Haagerup fusion ring, and define $\zeta \equiv {3 + \sqrt{13} \over 2}$.  For a unitary fusion category, the $F$-symbols involving at least one invertible object can be set to
\ie
\label{FSymbols}
(F^{\eta\cL\theta, \overline{\theta}\cL, \cL}_{\eta\cL})_{\eta, \overline\theta}
\ &= \
\zeta^{-1} \, ,
\qquad
(F^{\cL_1, \cL_3, \eta\cL_3}_{\cL_1\overline\eta})_{\cL_2, \overline\eta}
\ = \ 
(F^{\cL_1, \eta\cL_1, \eta\cL_3}_{\cL_3})_{\overline\eta, \cL_2}
\ = \ 
\zeta^{-\frac12} \, ,
\fe
where $\eta, \theta \in I$ and $\cL_i \in N$. The remaining nontrivial $F$-symbols are the ones where all six simple objects are non-invertible. It suffices to specify the nine components $(F^{\rho,\rho,\rho}_{*})_{\rho, *}$ with $*$ running over the non-invertible simple objects, since via the transparency relations
\ie
\label{FRelations1}
(F_{\cL_4}^{\cL_1, \cL_2, \cL_3})_{\cL_5, \cL_6}
\ &= \ (F_{\eta\cL_4}^{\eta\cL_1, \eta\cL_2, \eta\cL_3})_{\eta\cL_5, \eta\cL_6}
~=~
(F_{\cL_4}^{\eta\cL_1, \cL_2, \cL_3\eta})_{\cL_5, \cL_6}
\\
\ &= \ 
(F_{\cL_4\eta}^{\cL_1, \eta\cL_2, \cL_3})_{\cL_5, \cL_6}
~=~
(F_{\cL_4}^{\cL_1, \cL_2, \cL_3})_{\eta\cL_5, \cL_6\eta} \, 
\fe
the values of all other $F$-symbols are determined. An equivalent, and sometimes more convenient, set of relations that also allows the generation of all $F$-symbols is
\ie
\label{FRelations2}
(F^{\cL_1, \cL_2, \cL_3}_{\cL_4})_{\cL_5, \cL_6} \ &= \ (F^{\cL_1\overline\eta, \eta\cL_2, \cL_3}_{\cL_4})_{\cL_5, \eta\cL_6} \ 
= \ (F^{\cL_1, \cL_2\overline\eta, \eta\cL_3}_{\cL_4})_{\cL_5\overline\eta, \cL_6} \
\\
&= \ (F^{\cL_1, \cL_2, \cL_3\overline\eta}_{\cL_4\overline\eta})_{\cL_5, \cL_6\overline\eta} \
= \ (F^{\eta\cL_1, \cL_2, \cL_3}_{\eta\cL_4})_{\eta\cL_5, \cL_6} \, .
\fe
We provide an explicit algorithm to turn any nontrivial $F$-symbol into $(F^{\rho,\rho,\rho}_{*})_{\rho, *}$ form.
\begin{enumerate}
\item Use $(F_{\cL_4}^{\cL_1, \cL_2, \cL_3})_{\cL_5, \cL_6} = (F_{\cL_4}^{\eta\cL_1, \cL_2, \cL_3\eta})_{\cL_5, \cL_6}$ to turn $\cL_1$ into $\rho$.
\item Use $(F_{\cL_4}^{\cL_1, \cL_2, \cL_3})_{\cL_5, \cL_6} = (F_{\cL_4\eta}^{\cL_1, \eta\cL_2, \cL_3})_{\cL_5, \cL_6}$ to turn $\cL_2$ into $\rho$.
\item Use $(F^{\cL_1, \cL_2, \cL_3}_{\cL_4})_{\cL_5, \cL_6} = (F^{\cL_1, \cL_2, \cL_3\overline\eta}_{\cL_4\overline\eta})_{\cL_5, \cL_6\overline\eta}$ to turn $\cL_3$ into $\rho$.
\item Use $(F_{\cL_4}^{\cL_1, \cL_2, \cL_3})_{\cL_5, \cL_6} = (F_{\cL_4}^{\cL_1, \cL_2, \cL_3})_{\eta\cL_5, \cL_6\eta}$ to turn $\cL_5$ into $\rho$.
\end{enumerate}

Solving the pentagon identity in the gauge \eqref{FSymbols} and under transparency and $S_4$ full tetrahedral symmetry, there are exactly two solutions, both of which are unitary.  One of them is given by
\ie
\begin{array}{c|ccc}
(F^{\rho,\rho,\rho}_{*})_{\rho, *} & \rho & \A\rho & \A^2\rho
\\\hline
\rho & x & y_1 & y_2
\\
\A\rho & y_1 & y_2 & z
\\
\A^2\rho & y_2 & z & y_1
\end{array}
\fe
where
\ie
& x = {2 - \sqrt{13} \over 3} \, , \quad y_{1, 2} = \frac{1}{12} \left(5-\sqrt{13} \mp \sqrt{6 \left(1+\sqrt{13}\right)}\right) \, , \quad z = {1 + \sqrt{13} \over 6} \, .
\fe
The other solution is related by the $\text{Aut}(\bZ_3) \cong \bZ_2$ action that exchanges $y_1$ and $y_2$. For the first solution, some of the $F$-symbols can be presented as
\ie
& F^{\rho_i, \rho_j, \rho_j}_{\rho_i} =
\left(
\begin{array}{cccc}
\zeta^{-1} & \zeta^{-\frac12} & \zeta^{-\frac12} & \zeta^{-\frac12}
\\
\zeta^{-\frac12} & \kf_{i+j} & \kf_{i+j-1} & \kf_{i+j-2}
\\
\zeta^{-\frac12} & \kf_{i+j-1} & \kf_{i+j-2} & \kf_{i+j}
\\
\zeta^{-\frac12} & \kf_{i+j-2} & \kf_{i+j} & \kf_{i+j-1}
\end{array}
\right) \, ,
\qquad\quad
\kf_0 = x \, , \quad \kf_1 = y_2 \, , \quad \kf_2 = y_1 \, , 
\\
& F^{\rho_i,\rho_j,\rho_j}_{\rho_{k \neq i}} =
\left(
\begin{array}{ccc}
\kf'_{j-i-k} & \kf'_{j-i-k-1} & \kf'_{j-i-k-2}
\\
\kf'_{j-i-k-1} & \kf'_{j-i-k-2} & \kf'_{j-i-k}
\\
\kf'_{j-i-k-2} & \kf'_{j-i-k} & \kf'_{j-i-k-1}
\end{array}
\right) \, ,
\qquad
\kf'_0 = z \, , \quad \kf'_1 = y_2 \, , \quad \kf'_2 = y_1 \, ,
\\
& F^{\rho_i, \rho_j, \rho_i}_{\rho_{j \neq i}} = 
\left(
\begin{array}{ccc}
\kf'_{i+j} & \kf'_{i+j-1} & \kf'_{i+j-2}
\\
\kf'_{i+j-1} & \kf'_{i+j-2} & \kf'_{i+j}
\\
\kf'_{i+j-2} & \kf'_{i+j} & \kf'_{i+j-1}
\end{array}
\right) \, ,
\fe
where the subscripts of $\kf$ and $\kf'$ are defined modulo 3.

In this note we adopt a different, non-unitary, gauge to eliminate the appearance of some factors of $\zeta^{-\frac12}$ in the bootstrap equations.  The only difference from the unitary gauge is that \eqref{FSymbols} is replaced by
\ie
\label{FSymbolsAlt}
(F^{\eta\cL\theta,\overline{\cL\theta}, \cL}_{\eta\cL})_{\eta, \overline\theta}
~=~
(F^{\cL_1, \eta\cL_1, \eta\cL_3}_{\ocL_3})_{\overline\eta, \cL_2}
~=~
\zeta^{-1} \, ,
\qquad
(F^{\cL_1, \cL_3, \eta\cL_3}_{\overline{\eta\cL_1}})_{\cL_2, \overline\eta}
~=~
1 \, .
\fe
Consequently, the $4 \times 4$ $F$-symbols become
\ie
\label{FMatrix}
& F^{\rho_i, \rho_j, \rho_j}_{\rho_i} =
\left(
\begin{array}{cccc}
\zeta^{-1} & \zeta^{-1} & \zeta^{-1} & \zeta^{-1}
\\
1 & \kf_{i+j} & \kf_{i+j-1} & \kf_{i+j-2}
\\
1 & \kf_{i+j-1} & \kf_{i+j-2} & \kf_{i+j}
\\
1 & \kf_{i+j-2} & \kf_{i+j} & \kf_{i+j-1}
\end{array}
\right) \, ,
\fe
while the $3 \times 3$ $F$-symbols remain the same.

\section{Crossing symmetry of $\rho$ defect operators}
\label{App:Defect}

General crossing symmetry involving topological defect lines (TDLs) was discussed in Section~\ref{Sec:Crossing}.  In search for a defect topological field theory (TFT) whose TDLs  realize the Haagerup $\cH_3$ fusion category, the subset of crossing symmetry constraints that are equivalent to the associativity with at least one local operator was delineated in Section~\ref{Sec:Consistency}, and solved in Section~\ref{Sec:LocalAlgebra} to obtain part of the defining data of the TFT.  In this appendix, we study other crossing symmetry constraints that encode more data of the TFT.  These constraints can be depicted graphically as
\ie
\label{DefectCrossingDiagram}
\begin{gathered}
\begin{tikzpicture}[scale = 1]
\draw [line,-<-=.6] (-1,0) -- (-.5,0) node[above] {$\cL$} -- (0,0);
\draw [line] (-1,0) -- (-1.5,.87) \dt{above left}{-3}{$o_{i_1 A_1}$};
\draw [line] (-1,0) -- (-1.5,-.87) \dt{below left}{-3}{$o_{i_2 A_2}$};
\draw [line] (0,0) -- (.5,-.87) \dt{below right}{-3}{$o_{i_3 A_3}$};
\draw [line] (0,0) -- (.5,.87) \dt{above right}{-3}{$o_{i_4 A_4}$};
\draw [line,dotted] (-.2,-1.45) -- (-.2,1.45);
\end{tikzpicture}
\end{gathered}
~~ = \sum_{\cL'} ~~
\begin{gathered}
\begin{tikzpicture}[scale = 1]
\draw [line,-<-=.6] (0,-1) -- (0,-.5) node[right] {${\cL}'$} -- (0,0);
\draw [line] (0,-1) -- (.87,-1.5) \dt{below right}{-3}{$o_{i_3 A_3}$};
\draw [line] (0,-1) -- (-.87,-1.5) \dt{below left}{-3}{$o_{i_2 A_2}$};
\draw [line] (0,0) -- (-.87,.5) \dt{above left}{-3}{$o_{i_1 A_1}$};
\draw [line] (0,0) -- (.87,.5) \dt{above right}{-3}{$o_{i_4 A_4}$};
\draw [line,dotted] (-1.45,-.2) -- (1.45,-.2);
\end{tikzpicture}
\end{gathered} 
~~
(F^{\rho_{i_1}, \rho_{i_2}, \rho_{i_3}}_{\rho_{i_4}})_{\cL, \cL'}
\, ,
\fe
and cutting along the dotted line gives
\ie
\label{DefectCrossing}
& \sum_{\cO \in \cH_\cL} c(o_{i_1 A_1}, o_{i_2 A_2}, \cO) \, c(o_{i_3 A_3}, o_{i_4 A_4}, \ocO) 
\\
&= \sum_{\cL'} (F^{\rho_{i_1}, \rho_{i_2}, \rho_{i_3}}_{\rho_{i_4}})_{\cL, \cL'} \sum_{\cO' \in \cH_{\cL'}} c(o_{i_2 A_2}, o_{i_3 A_3}, \cO') \, c(o_{i_4 A_4}, o_{i_1 A_1}, \ocO') \, .
\fe
Depending on the quadruple $(i_1, i_2, i_3, i_4)$, the internal TDLs $\cL, \, \cL'$ run over either the three non-invertible TDLs $\rho_0 \equiv \rho, ~ \rho_1 \equiv \A\rho, ~ \rho_2 \equiv \A^2\rho$, or an additional invertible TDL.  It is convenient to introduce a capital $I$ index such that $\rho_{I = -1}$ denotes the invertible TDL whenever applicable, and $\rho_{I = i} = \rho_i$ for $i = 0, 1, 2$.  In particular, if $\rho_{I = -1}$ is the trivial TDL $\cI$, then $o_{I=-1, A}$ with $A = 1, \dotsc, n_{\rm V}$ represent the local operators.

\subsubsection*{Two pairs of identical external operators $o_{iA}$ and $o_{jB}$ in the 1221 configuration}

In this case, the defect crossing equation \eqref{DefectCrossing} in the newly introduced notation reads
\ie
&
\begin{gathered}
\begin{tikzpicture}[scale = 1]
\draw [line,-<-=.6] (-1,0) -- (-.5,0) node[above] {$\rho_K$} -- (0,0);
\draw [line] (-1,0) -- (-1.5,.87) \dt{above left}{-3}{$o_{i A}$};
\draw [line] (-1,0) -- (-1.5,-.87) \dt{below left}{-3}{$o_{j B}$};
\draw [line] (0,0) -- (.5,-.87) \dt{below right}{-3}{$o_{j B}$};
\draw [line] (0,0) -- (.5,.87) \dt{above right}{-3}{$o_{i A}$};
\draw [line,dotted] (-.2,-1.45) node[below]{$|o_{KC}\rangle \langle o_{KC}|$} -- (-.2,1.45) node[above]{~};
\end{tikzpicture}
\end{gathered}
~~ = \sum_{L=-1}^2 ~~
\begin{gathered}
\begin{tikzpicture}[scale = 1]
\draw [line,-<-=.6] (0,-1) -- (0,-.5) node[right] {$\rho_L$} -- (0,0);
\draw [line] (0,-1) -- (.87,-1.5) \dt{below right}{-3}{$o_{j B}$};
\draw [line] (0,-1) -- (-.87,-1.5) \dt{below left}{-3}{$o_{j B}$};
\draw [line] (0,0) -- (-.87,.5) \dt{above left}{-3}{$o_{i A}$};
\draw [line] (0,0) -- (.87,.5) \dt{above right}{-3}{$o_{i A}$};
\draw [line,dotted] (-1.45,-.2) 
-- (1.45,-.2) node[right]{$|o_{LD}\rangle \langle o_{LD}|$};
\end{tikzpicture}
\end{gathered} 
~~
(F^{\rho_{i_1}, \rho_{i_2}, \rho_{i_3}}_{\rho_{i_4}})_{\rho_K, \rho_L}
\, ,
\\
\label{ijji}
& \sum_{C=-1}^2 |c(o_{iA}, o_{jB}, o_{KC})|^2
=
\sum_{L=-1}^2 (F^{\rho_i, \rho_j, \rho_j}_{\rho_i})_{\rho_K, \rho_L} \, \sum_D c(o_{iA}, o_{iA}, o_{LD}) \, c(o_{jB}, o_{jB}, o_{LD}) \, .
\fe

\begin{itemize}

\item Setting $K = -1$ gives
\ie
\label{PreSelection}
\sum_C c(o_{iA}, o_{jB}, o_{-1,C})^2 &= \zeta^{-1} \sum_D c(o_{iA}, o_{iA}, o_{-1,D}) \, c(o_{jB}, o_{jB}, o_{-1,D})
\\
&+ \zeta^{-1} \sum_{\ell=0}^2 \sum_D c(o_{iA}, o_{iA}, o_{\ell D}) \, c(o_{jB}, o_{jB}, o_{\ell D}) \, ,
\fe
where we have used the explicit values of $F$-symbols given in \eqref{FMatrix}.  

\item
If we sum \eqref{ijji} over $K = k = 0, 1, 2$ (but not $K = -1$), 
and use the explicit values of $F$-symbols given in \eqref{FMatrix}, then we obtain\footnote{We used
\ie
& \sum_{k=0}^2 (F^{\rho_i, \rho_j, \rho_j}_{\rho_i})_{\rho_k, \cI} = 3 \, ,
\qquad
\sum_{k=0}^2 (F^{\rho_i, \rho_j, \rho_j}_{\rho_i})_{\rho_k, \rho_\ell} = x + y_+ + y_- = - \zeta^{-1} \quad \forall \, \ell \, .
\fe
}
\ie
\label{SumK}
\sum_{k=0}^2 \sum_C |c(o_{iA}, o_{jB}, o_{kC})|^2 &= 3 \sum_D c(o_{iA}, o_{iA}, o_{-1,D}) \, c(o_{jB}, o_{jB}, o_{-1,D})
\\
&- \zeta^{-1} \sum_{\ell=0}^2 \sum_D c(o_{iA}, o_{iA}, o_{\ell D}) \, c(o_{jB}, o_{jB}, o_{\ell D}) \, .
\fe
\item
Let us set $i = j$ and $A \neq B$.  Using the explicit values of $\kappa^i_{AB}$ and $\lambda^i_{AB;a}$ in \eqref{kappaEigvals} and \eqref{lambda} to evaluate the contributions from local operators,
\ie
\sum_C c(o_{iA}, o_{iB}, o_{-1,C})^2 &= \D_{AB} + (\kappa^i_{AB})^2 + 2 \sum_a \lambda^i_{AB;a} \bar\lambda^i_{AB;a}
=
\begin{cases}
3 \sigma_{111}^2
& A = B \, ,
\\
0 & A \neq B \, ,
\end{cases}
\fe
\ie
&\hspace{-.1in} \sum_D c(o_{iA}, o_{iA}, o_{-1,D}) \, c(o_{jB}, o_{jB}, o_{-1,D}) 
\\
&= 1 + \kappa^i_{AA} \kappa^j_{BB} + \sum_a \left( \lambda^i_{AA;a} \bar\lambda^j_{BB;a} + \bar\lambda^i_{AA;a} \lambda^j_{BB;a} \right)
=
\begin{cases}
3 \sigma_{111}^2 & A - i = B - j \, ,
\\
0 & A - i \neq B - j \, ,
\end{cases}
\fe
where
\ie
\sigma_{111} = \sqrt{1 + \zeta^{-2}} \, ,
\fe
the preceding two equations \eqref{PreSelection} and \eqref{SumK} become
\ie
\label{AABB}
0 &= \sum_{\ell=0}^2 \sum_D c(o_{iA}, o_{iA}, o_{\ell D}) \, c(o_{iB}, o_{iB}, o_{\ell D}) \, ,
\\
\sum_{k=0}^2 \sum_C c(o_{iA}, o_{iB}, o_{kC})^2 &= - \zeta^{-1} \sum_{\ell=0}^2 \sum_D c(o_{iA}, o_{iA}, o_{\ell D}) \, c(o_{iB}, o_{iB}, o_{\ell D}) \, .
\fe
It follows that
\ie
\sum_{k=0}^2 \sum_C c(o_{iA}, o_{iB}, o_{kC})^2 = 0 \, ,
\fe
and we arrive at a selection rule:
\ie
\label{SelectionRule}
c(o_{iA}, o_{iB}, o_{kC}) = 0 \quad \forall \, i, \, k, \, C \quad \text{if $A \neq B$}.
\fe

\end{itemize}

\subsubsection*{Four identical external defect operators $o_{iA}$}

In this case, the defect crossing equation \eqref{DefectCrossing} becomes
\ie
&
\begin{gathered}
\begin{tikzpicture}[scale = 1]
\draw [line,-<-=.6] (-1,0) -- (-.5,0) node[above] {$\rho_J$} -- (0,0);
\draw [line] (-1,0) -- (-1.5,.87) \dt{above left}{-3}{$o_{i A}$};
\draw [line] (-1,0) -- (-1.5,-.87) \dt{below left}{-3}{$o_{i A}$};
\draw [line] (0,0) -- (.5,-.87) \dt{below right}{-3}{$o_{i A}$};
\draw [line] (0,0) -- (.5,.87) \dt{above right}{-3}{$o_{i A}$};
\draw [line,dotted] (-.2,-1.45) node[below]{$|o_{JB}\rangle \langle o_{JB}|$} -- (-.2,1.45) node[above]{~};
\end{tikzpicture}
\end{gathered}
~~ = \sum_{K=-1}^2 ~~
\begin{gathered}
\begin{tikzpicture}[scale = 1]
\draw [line,-<-=.6] (0,-1) -- (0,-.5) node[right] {$\rho_K$} -- (0,0);
\draw [line] (0,-1) -- (.87,-1.5) \dt{below right}{-3}{$o_{i A}$};
\draw [line] (0,-1) -- (-.87,-1.5) \dt{below left}{-3}{$o_{i A}$};
\draw [line] (0,0) -- (-.87,.5) \dt{above left}{-3}{$o_{i A}$};
\draw [line] (0,0) -- (.87,.5) \dt{above right}{-3}{$o_{i A}$};
\draw [line,dotted] (-1.45,-.2) 
-- (1.45,-.2) node[right]{$|o_{KC}\rangle \langle o_{KC}|$};
\end{tikzpicture}
\end{gathered} 
~~
(F^{\rho_{i_1}, \rho_{i_2}, \rho_{i_3}}_{\rho_{i_4}})_{\rho_J, \rho_K}
\, ,
\\
& \sum_{K=-1}^2 \left[ \D_{JK} \D_{BC} - (F^{\rho_i, \rho_i, \rho_i}_{\rho_i})_{\rho_J, \rho_K} \right] \, \sum_C c(o_{iA}, o_{iA}, o_{KC})^2 = 0 \, ,
\fe
which says that the four-dimensional vector $\sum_C c(o_{iA}, o_{iA}, o_{KC})^2$ is a non-negative four-dimensional eigenvector of the matrix $F^{\rho_i, \rho_i, \rho_i}_{\rho_i}$ with eigenvalue one.  Using the explicit values of the $F$-symbols given in \eqref{FMatrix}, we determine that such an eigenvector is in the two-dimensional subspace spanned by
\ie
\left( 1+\zeta, \, 1, \, 1, \, 1 \right) \, ,
\quad
\begin{cases}
\left( 0, \, 1, \, \psi_+, \, \psi_- \right) & i = 0 \, ,
\\
\left( 0, \, \psi_-, \, 1, \, \psi_+ \right) & i = 1 \, ,
\\
\left( 0, \, \psi_+, \, \psi_-, \, 1 \right) & i = 2 \, ,
\end{cases}
\fe
where
\ie
\psi_\pm = {-1 \pm \sqrt{7 + 2\sqrt{13}} \over 2} \, .
\fe
Equivalently, $\sum_C c(o_{iA}, o_{iA}, o_{KC})^2$ is orthogonal to 
\ie
(-1+\zeta^{-1}, \, 1, \, 1, \, 1) \, ,
\quad 
\begin{cases}
\left( 0, \, 1, \, \eta_-, \, \eta_+ \right) & i = 0 \, ,
\\
\left( 0, \, \eta_+, \, 1, \, \eta_- \right) & i = 1 \, ,
\\
\left( 0, \, \eta_-, \, \eta_+, \, 1 \right) & i = 2 \, ,
\end{cases}
\fe
where
\ie
\eta_\pm = {-1\pm\sqrt{3 \left(2 \sqrt{13}-7\right)} \over 2} \, .
\fe

\subsubsection*{Two pairs of identical external operators $o_{iA}$ and $o_{jB}$ in the 1212 configuration}

We first recall from \eqref{InvCyclic} and \eqref{ConjReflect} that the defect three-point coefficients are invariant under cyclic permutations and complex conjugate under reflections.  Thus the trivalent vertices 
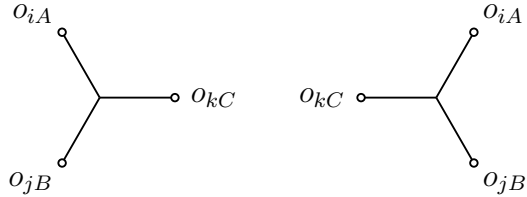
\begin{center}
\begin{tikzpicture}
\draw [line] (0,0) -- (1,0) \dt{right}{0}{$o_{kC}$};
\draw [line] (0,0) -- (-.5,.87) \dt{above left}{-3}{$o_{iA}$};
\draw [line] (0,0) -- (-.5,-.87) \dt{below left}{-3}{$o_{jB}$};
\end{tikzpicture}
\quad
\begin{tikzpicture}
\draw [line] (0,0) -- (-1,0) \dt{left}{0}{$o_{kC}$};
\draw [line] (0,0) -- (.5,-.87) \dt{below right}{-3}{$o_{jB}$};
\draw [line] (0,0) -- (.5,.87) \dt{above right}{-3}{$o_{iA}$};
\end{tikzpicture}
\end{center}
are complex conjugates of each other, and can differ by a phase $2\phi_{ijk}$.  The corresponding three-point coefficients of defect operators can be parameterized as
\ie
\label{Phase}
& c(o_{iA}, o_{jB}, o_{kC}) = |c(o_{iA}, o_{jB}, o_{kC})| e^{i\phi_{ijk}} \, ,
\\
& c(o_{jB}, o_{iA}, o_{kC}) = |c(o_{iA}, o_{jB}, o_{kC})| e^{-i\phi_{ijk}}  \, .
\fe
Since the phase is trivial when two indices coincide, $\phi_{iij} = 0$, the only nontrivial phase is $\phi \equiv \phi_{012}$.  Let us define
\ie
\Phi_{ij} = \begin{pmatrix}
e^{i\phi_{ij0}} 
\\
& e^{i\phi_{ij1}}
\\
&& e^{i\phi_{ij2}}
\end{pmatrix} \, ,
\fe
which is the identity matrix if $i = j$, and has a single possibly nontrivial entry if $i \neq j$.  

In the current case, the crossing equation \eqref{DefectCrossing} becomes
\ie
&
\begin{gathered}
\begin{tikzpicture}[scale = 1]
\draw [line,-<-=.6] (-1,0) -- (-.5,0) node[above] {$\rho_k$} -- (0,0);
\draw [line] (-1,0) -- (-1.5,.87) \dt{above left}{-3}{$o_{i A}$};
\draw [line] (-1,0) -- (-1.5,-.87) \dt{below left}{-3}{$o_{j B}$};
\draw [line] (0,0) -- (.5,-.87) \dt{below right}{-3}{$o_{i A}$};
\draw [line] (0,0) -- (.5,.87) \dt{above right}{-3}{$o_{j B}$};
\draw [line,dotted] (-.2,-1.45) node[below]{$|o_{kC}\rangle \langle o_{kC}|$} -- (-.2,1.45) node[above]{~};
\end{tikzpicture}
\end{gathered}
~~ = \sum_{\ell=0}^2 ~~
\begin{gathered}
\begin{tikzpicture}[scale = 1]
\draw [line,-<-=.6] (0,-1) -- (0,-.5) node[right] {$\rho_L$} -- (0,0);
\draw [line] (0,-1) -- (.87,-1.5) \dt{below right}{-3}{$o_{i A}$};
\draw [line] (0,-1) -- (-.87,-1.5) \dt{below left}{-3}{$o_{j B}$};
\draw [line] (0,0) -- (-.87,.5) \dt{above left}{-3}{$o_{i A}$};
\draw [line] (0,0) -- (.87,.5) \dt{above right}{-3}{$o_{j B}$};
\draw [line,dotted] (-1.45,-.2) -- (1.45,-.2) node[right]{$|o_{\ell D}\rangle \langle o_{\ell D}|$};
\end{tikzpicture}
\end{gathered} 
~~
(F^{\rho_{i_1}, \rho_{i_2}, \rho_{i_3}}_{\rho_{i_4}})_{\rho_k, \rho_\ell}
\, ,
\\
&\sum_C c(o_{iA}, o_{jB}, o_{kC})^2 = \sum_\ell (F^{\rho_i, \rho_j, \rho_i}_{\rho_j})_{\rho_k, \rho_\ell} \, \sum_D c(o_{jB}, o_{iA}, o_{\ell D})^2 \, .
\fe
By factoring out the phase using \eqref{Phase}, the above crossing equation can be reexpressed as
\ie
\label{ijij}
\sum_C |c(o_{iA}, o_{jB}, o_{kC})|^2 = \sum_{\ell=0}^2 
(\bar\Phi_{ij}^2 \, F^{\rho_i, \rho_j, \rho_i}_{\rho_j} \, \bar\Phi_{ij}^2)_{k\ell} \, 
\sum_D |c(o_{iA}, o_{jB}, o_{\ell D})|^2 \, .
\fe
The three-dimensional vector $\sum_C |c(o_{iA}, o_{jB}, o_{kC})|^2$, if nonzero, is a non-negative eigenvector of $\bar\Phi_{ij}^2 \, F^{\rho_i, \rho_j, \rho_i}_{\rho_j} \, \bar\Phi_{ij}^2$ with eigenvalue 1.

\begin{itemize}

\item
If $\phi$ is a generic phase, by which we mean $\phi \neq 0, \, \pi$, then such an eigenvector is unique up to overall normalization, given by
\ie
\begin{cases}
\left( \psi_+ , \, - \psi_-, \, 0 \right) \quad \{i, j\} = \{0, 1\} \, ,
\\
\left( 0 , \, \psi_+ , \, -\psi_- \right) \quad \{i, j\} = \{1, 2\} \, ,
\\
\left( - \psi_- , \, 0, \, \psi_+ \right) \quad \{i, j\} = \{0,2\} \, ,
\end{cases}
\fe
which in particular implies that
\ie
c(o_{0A}, o_{1B}, o_{2C}) = 0 \, .
\fe
In other words, the only three-point coefficient that is allowed to have a nontrivial phase vanishes.  Then without loss of generality, we can assume $\phi = 0, \, \pi$.

\item
If $\phi = 0, \, \pi$, then the eigenvector is in the two-dimensional subspace that is orthogonal to
\ie
\begin{cases}
\left( \psi_- , \, \psi_+, \, e^{i\phi} \right) \quad \{i, j\} = \{0, 1\} \, ,
\\
\left( e^{i\phi} , \, \psi_- , \, \psi_+ \right) \quad \{i, j\} = \{1, 2\} \, ,
\\
\left( \psi_+ , \, e^{i\phi}, \, \psi_- \right) \quad \{i, j\} = \{0,2\} \, .
\end{cases}
\fe

\end{itemize}

\bibliography{refs} 
\bibliographystyle{JHEP}

\end{document}